# A Review on Improving PSC Performance through Charge Carrier Management: Where We Stand and What's Next?

## Dilshod Nematov [1,2] *

[1]S.U. Umarov Physical-Technical Institute of NAST, Dushanbe 734063, Tajikistan

[2]School of Optoelectronic Engineering & CQUPT-BUL Innovation Institute, Chongqing University of Posts and Telecommunications, Chongqing 400065, China

* Corresponding author: dilnem@mail.ru

**Abstract** - Perovskite solar cells (PSCs) represent a breakthrough in photovoltaic technology, combining high power conversion efficiencies (PCEs), ease of fabrication, and tunable optoelectronic properties. However, their commercial viability is limited by critical issues such as charge carrier recombination, interfacial defects, instability under environmental stress, and toxicity of lead-based components. This review systematically examines recent advancements in charge carrier management strategies aimed at overcoming these limitations. Initially, fundamental mechanisms governing carrier generation, separation, transport, and recombination are outlined to provide a clear foundation. The study then delves into an in-depth analysis of carrier lifetime and mobility, evaluating recent methodologies for their enhancement through compositional engineering and structural optimization. Subsequently, trap state passivation techniques and interface engineering approaches are reviewed, with a particular focus on their impact on device stability and efficiency. The review also discusses long-term stability strategies and emerging trends in lead-free and scalable PSC technologies. In this work, recent strategies for charge carrier management are systematically categorized, comparative analyses are provided and synergistic solutions with high potential for real-world implementation are highlighted. By synthesizing data and perspectives from over thirty recent studies, this article offers a comprehensive roadmap for researchers seeking to optimize PSC performance and accelerate their transition toward commercial application.

## 1. Introduction

Over the past decade, perovskite solar cells (PSCs) have made unprecedented advances in the field of photovoltaics, achieving a power conversion efficiency (PCE) of over 26% for single-junction devices [1] and more than 33% in tandem configurations with silicon [2]. These achievements have been made possible due to the unique properties of hybrid perovskites, including strong light absorption, tunable bandgap and exceptional charge carrier diffusion lengths [3]. However, the transition from laboratory-scale prototypes to commercial modules faces critical challenges such as degradation caused by light, heat and moisture [4], lead toxicity [5], and instability in device performance as the active area increases [6]. A key factor underlying these issues is the management of charge carriers, encompassing their generation, transport and the suppression of recombination losses, which collectively determine both the efficiency and long-term operational stability of PSCs [7].

Despite significant progress, a major issue remains unresolved. The high PCE observed in small-area devices tends to drop sharply when scaling up to modules with areas exceeding

100 cm$^2$ [10]. This decline is primarily associated with morphological inhomogeneity, charge accumulation at grain boundaries [9], and an imbalance in the mobilities of electrons and holes [14], which is further aggravated by halide ion migration [8]. Conventional approaches, such as defect passivation using 2D perovskites [11] or Rb$^+$ doping [12], have shown limited effectiveness under practical operating conditions, typically 85 $^o$C and 85% relative humidity [13]. Moreover, most studies address these challenges in isolation and often overlook the complex interrelations between microstructure, charge kinetics and degradation pathways.

This article proposes an interdisciplinary strategy for charge carrier management by integrating recent developments in materials science, interface engineering and computational modeling. Emerging evidence highlights critical interdependencies. For example, both crystal orientation and grain size directly influence recombination behavior [16], while ion migration redistributes internal electric fields and disrupts charge transport [17]. Novel approaches, such as covalent organic frameworks [18] and ionic liquid additives [19], offer promising routes not only for improving stability but also for enabling precise control over carrier dynamics.

Particular attention is given to the challenges associated with scaling up. The transfer of laboratory-scale strategies, such as atomically precise grain boundary passivation [20], to industrial modules exceeding 400 cm$^2$ [21] is complicated by thermomechanical stress and coating non-uniformity, both of which contribute to further imbalance in charge transport. Additionally, the need to balance efficiency with environmental safety calls for alternatives to lead, such as tin and germanium [22]. However, these substitutes tend to reduce carrier diffusion lengths, necessitating a reconsideration of charge transport layer design.

This review is organized into the following sections: (1) fundamental mechanisms of charge generation, transport and recombination; (2) strategies for enhancing carrier lifetime and mobility; (3) approaches for minimizing recombination losses; (4) interface engineering and morphological control; and (5) commercial prospects for PSC technology. Unlike previous reviews, this work incorporates environmental concerns, such as lead leaching [23] and module recycling [24, 25], into the broader context of charge carrier management. It also highlights breakthrough technologies, including hybrid tandem architectures [27], polymer matrix encapsulation [23, 26], and the application of machine learning techniques for optimizing morphology and defect distribution [28]. These directions not only broaden the potential of PSCs but also help to lay the foundation for a more sustainable solar energy future.

## 2. Fundamentals of Charge Carrier Dynamics in PSCs

The performance of perovskite solar cells (PSCs) is governed by a sequence of interrelated processes, including the photogeneration of charge carriers, their transport through the active and transport layers, and recombination. These processes are highly sensitive to the morphology and crystallographic structure of the active layer, as well as to the nature of grain boundaries, interfaces, and defects. A deep understanding of these mechanisms is essential for improving both efficiency and long-term stability.

Due to their direct bandgap and high absorption coefficient (approximately 10$^5$ cm$^{-1}$), perovskite materials are capable of harvesting sunlight effectively, even in thin layers ranging from 200 to 500 nm. Their absorption spectrum shows that a film thickness above 300 nm is sufficient for nearly complete absorption of visible light [16, 29]. This makes it possible to reduce the amount of active material used and to improve overall device architecture.

In PSCs, electron-hole pairs are generated primarily within the perovskite layer. The diffusion length of these carriers varies depending on the material quality. In single-crystal perovskites, diffusion lengths can reach 10 to 20 µm. In contrast, in polycrystalline films, which are widely used in both laboratory and commercial devices, this length is typically between 0.5 and 3 µm. The presence of grain boundaries in such films acts as a source of trap states and facilitates ion migration. The morphology of the perovskite layer, especially the grain size and crystallographic orientation, plays a critical role in determining the efficiency of charge transport [21, 30]. Crystal orientation can introduce strong anisotropy in carrier mobility. Spectroscopic data and first-principles calculations indicate that alignment along the (110) plane supports greater carrier delocalization and leads to higher mobility compared to the (100) orientation [16]. This has been confirmed by fluorescence mapping and time-resolved photoluminescence (TRPL) imaging.

Along with the grain size, the passivation of surface and interface defects is also of great importance. Bulk trap states caused by vacancies and interstitials, particularly iodine ($I^-$) and lead ($Pb^{2+}$) defects, are known to promote non-radiative recombination through the Shockley–Read–Hall (SRH) mechanism. When such traps are present, carrier lifetimes can drop to a few hundred nanoseconds [31]. Surface treatment with fluorinated salts or the incorporation of self-assembled monolayers (SAMs) has been shown to improve carrier lifetimes and reduce leakage currents. These effects have been confirmed by current–voltage (J-V) characteristics and TRPL measurements [32, 33].

Interfaces between the perovskite layer and the electron and hole transport layers (ETL and HTL) are also critical. Poor interface quality can create non-radiative recombination pathways and lead to the formation of high-density surface states, reaching values up to $10^{17}$ cm$^{-3}$. Passivation of ETLs and HTLs using organic and inorganic interlayers such as NiO, $C_{60}$, and PEAI has been reported to improve both the open-circuit voltage ($V_{oc}$) and the overall device efficiency [11]. Ion migration is another factor that contributes significantly to PSC instability and degradation. Under thermal stress and applied bias, mobile ions, including $I^-$, $Br^-$, and $MA^+$, can move along grain boundaries, leading to local charge imbalances and the formation of internal electric fields that accelerate recombination [18–23]. The use of larger organic cations, such as guanidinium or phenylethylammonium, as well as 2D perovskite layers, has been shown to reduce ion mobility and improve stability [26].

The choice of charge transport materials also influences overall device performance. $TiO_2$, a widely used ETL material, suffers from photocatalytic activity and charge accumulation, which limits long-term stability. $SnO_2$ is a promising alternative due to its higher electron mobility and better energetic alignment with the perovskite absorber [34–36]. Improved current–voltage characteristics and external quantum efficiency (EQE) have been observed in devices employing $SnO_2$, particularly when doped with cesium or aluminum [34, 35]. On the hole transport side, replacing the commonly used Spiro-OMeTAD with inorganic materials such as NiO or CuSCN offers better thermal stability and reduced cost. These materials are less sensitive to environmental degradation and do not require complex dopants. A holistic approach to optimizing PSC performance must consider all of these factors, including perovskite morphology, defect passivation, suppression of ion migration, and the selection of efficient ETL and HTL materials. Integrating these improvements leads to significant gains in photovoltaic performance.

Another important aspect is the ultrafast photoinduced dynamics of charge carriers, which directly affect charge separation, recombination rates, and ultimately the efficiency of the device. Figure 1 illustrates the main photoinduced processes in metal halide perovskites, each occurring on a characteristic timescale [37]. Upon photon absorption (step 1), electron–hole pairs are generated in the perovskite layer. This is followed by carrier thermalization (step 2), which occurs within femtoseconds. Subsequently, some carriers become trapped in defect states (step 3), while others recombine through band-to-band transitions (step 4) or diffuse toward the interfaces (step 5). Carriers that reach the interface are extracted by the transport layers (step 6). Those trapped at defect sites may recombine through localized states (step 7), reducing the quantum efficiency of the device.

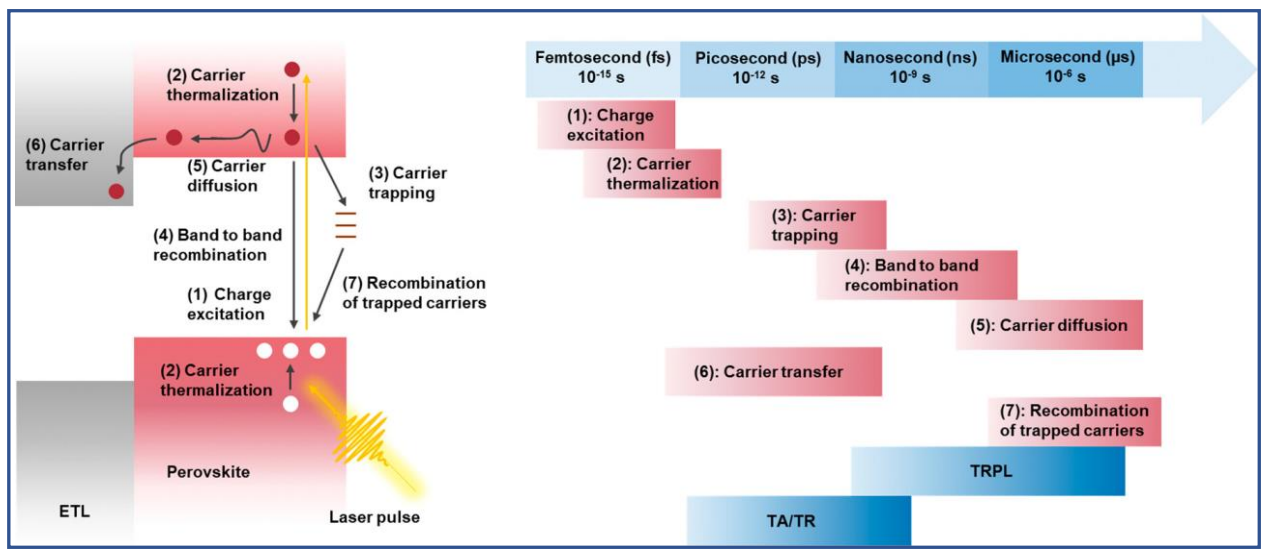

**Figure 1.** Schematic representation of the complex photophysical processes occurring in the charge transport layer/perovskite stack following photoexcitation, along with the characteristic timescales of each process [37].

Metal halide perovskites, except for their two-dimensional forms, exhibit low exciton binding energies in the range of approximately 10 to 50 meV. This property allows them to efficiently dissociate into free charge carriers at room temperature immediately after light absorption. When photons possess energy greater than the bandgap, high-energy carriers, also known as hot carriers, are generated. These carriers undergo thermalization, which is the process of energy redistribution between carriers and the lattice, and this takes place within about 100 femtoseconds [38]. After that, the carriers cool and relax toward the edges of the conduction and valence bands over a timescale of approximately 1 picosecond [39]. Since thermal relaxation results in the loss of excess energy as heat, many studies have focused on extracting hot carriers before they undergo full thermalization. This approach has the potential to increase device efficiency beyond the Shockley–Queisser limit [40].

Following thermalization, charge carriers may either recombine radiatively, accompanied by photon emission, or become trapped at defect sites where non-radiative recombination occurs. These events typically take place on timescales ranging from nanoseconds to several microseconds, as revealed by time-resolved photoluminescence (TRPL) studies [41]. Early-stage recombination through surface or grain boundary defects is common in unpassivated films. In some cases, trapped carriers may participate in photodoping, which leads to long-term changes in material properties [42]. Monitoring these ultrafast processes is possible using time-resolved spectroscopy techniques such as transient absorption and photoluminescence. These methods help identify the lifetimes associated with carrier generation, relaxation, and recombination [42].

A detailed understanding of these mechanisms enables better control over charge dynamics, ultimately supporting improvements in both efficiency and device stability.

The portion of carriers that avoid recombination and trapping may reach the interface between the perovskite absorber and the charge transport layer. Their movement is driven by energy level offsets between the perovskite and the transport material, which create an internal electrochemical field that promotes directional charge flow [43–45]. In practical devices, however, charge trapping at interfacial defects or within bulk trap states may compete with transport. This makes it difficult to isolate the influence of individual processes on the overall carrier dynamics. Defects located near the interface can form energy barriers that hinder effective carrier collection [46]. This situation becomes more critical in films that exhibit large gradients in crystallinity, high defect densities, or significant morphological inhomogeneity. In particular, grain boundaries may act as traps that limit carrier diffusion [47].

The kinetics of charge transport and extraction are also influenced by factors such as the thickness of the active layer, grain structure, and the method used to synthesize the perovskite film. Differences in processing techniques, including solvent engineering, vapor-assisted post-treatment, and annealing conditions, can lead to variations in defect density. As a result, carrier lifetimes and extraction rates may vary widely [48]. Studies have shown that even for the same perovskite composition, such as MAPbI$_3$, extraction times can differ by several orders of magnitude depending on morphology and interfacial engineering [49]. Representative examples illustrating such variations are summarized in Table 1. Careful control over interfacial quality, defect passivation, and perovskite morphology is essential for optimizing charge transport and improving both the efficiency and operational stability of perovskite solar cells.

**Table 1.** Reported range of charge transfer (CT) lifetimes (τ) in various CTL/PSC architectures.

| Method | Architec-ture | Meas. Param. | Evaluation method | $\tau_{CT}$ | Rate constant ($10^7$ s$^{-1}$) | Ref. |
|---|---|---|---|---|---|---|
| TA | Spiro-MeOTAD/ CH$_3$NH$_3$PbI$_3$ | λexc = 600 nm, 10.0 μJcm$^{-2}$ | multiexponential fitting, the short lifetime component is attributed to CT | 0.7 ns | 142.9 | 50 |
| | | λexc = 485 nm, 0.5−75 μJcm$^{-2}$ | global analysis of data, CT is determined from the fluence dependence of average lifetime | 50 ns | 2.0 | 51 |
| | | λexc = 370 nm, 3.0 μJcm$^{-2}$ | multiexponential fitting to lower wavelength than the ground state bleach | <10 ps | >10000 | 52 |
| | | λexc = 464 nm, unknown fluence | triexponential fitting, the fast component is attributed to CT | 2.1 ns | 47.6 | 53 |
| | | λexc = 485 nm, 0.5−75 μJcm$^{-2}$ | global analysis of data, CT is determined from the fluence dependence | 17 ns | 5.9 | 54 |
| | | | of average lifetime | | | |
| TRPL | TiO$_2$/CH$_3$NH$_3$PbI$_3$ | λexc=625 nm, 0.1 μJcm$^{-2}$ | CT equation was used with derived τeffective (Al$_2$O$_3$/perovskite acted as reference without CT) | 11 ns | 9.1 | 55 |
| | | λexc = 460 nm, 2.0 μJcm$^{-2}$ | multiexponential fitting | 0.8 ps | 125000 | 56 |
| | | λexc = 640 nm, unknown fluence | biexponential fitting, the fast component is attributed to CT | 9.7 ns | 10.3 | 57 |
| TRPL | PCBM/MAPbI$_3$ | λexc = 405 nm, 3 nJ cm$^{-2}$ | Bi-exponential fitting | 1.4 ns | 71.4 | 58 |
| | Spiro-MeOTAD/FAMA perovskite | λexc = 460 nm, 0.4 Wcm$^{-2}$ | global analysis of data, CT is determined from the fluence dependence of average lifetime | 100 ns | 1 | 59 |
| | | λexc = 625 nm, <0.1 μJcm$^{-2}$ | CT equation was used with derived τeffective (Al$_2$O$_3$/perovskite acted as reference without CT) | 1.8 ns | 55.6 | 55 |
| TA | Compact-TiO$_2$/CH$_3$NH$_3$PbI$_3$ | λexc = 400 nm, 10 μJcm$^{-2}$ | multiexponential fitting, the long lifetime component is attributed to CT | 370 ps | 270.3 | 60 |
| TRPL | Spiro-MeOTAD/ CH$_3$NH$_3$PbI$_3$ | λexc = 600 nm, 1.3 μJcm$^{-2}$ | CT equation was used with lifetimes determined from monoexponential fitting | 0.7 ns | 142.9 | 50 |
| TRPL | PCBM/MAPbI$_3$ | λexc = 600 nm, 1.3 μJ cm$^{-2}$ | Mono-exponential fitting | 0.4 ns | 250 | 50, 61 |
| TA | c-TiO$_2$/MAPbI$_3$ | λexc = 390 and 600 nm | Multiexponential fitting | 39.9 ps | 25.1 | 62 |
| TA | c-TiO$_2$/mp-TiO$_2$/MAPbI$_3$ | n/a | Multiexponential fitting | 150 ps | 6.7 | 62 |
| TA | mp-TiO$_2$/MAPbI$_3$ | λexc = 400 nm | Multiexponential fitting | 260–307 ps | 3.3–3.8 | 63 |
| TRPL | TiO$_2$/MAPbI$_3$ | λexc = 40 nm | Bi-exponential fitting | 2.3–5.8 ns | 17.2–43.5 | 64 |
| TRPL | TiO$_2$ single crystal/MAPbI$_3$ | λexc = 60 nm, 0.63 μJ cm$^{-2}$ | Bi-exponential fitting | 0.17–20.6 ns | 4.9–588.2 | 65 |
| TRPL | PCBM/MAPbI$_3$ | λexc = 635 nm, 0.06–0.13 mW cm$^{-2}$ | Bi-exponential fitting | 36 ns | 2.8 | 66 |
| TA | mp-TiO$_2$/graphene | λexc = 400 nm | Multiexponential fitting | 90–106 | 9.4–11.1 | 63 |

|  | QD/MAPbI$_3$ |  |  | ps |  |  |
| --- | --- | --- | --- | --- | --- | --- |
| TRPL | PCBM/MAPbI$_3$ | λexc = 464 nm | Tri-exponential fitting | 1.3 ns | 76.9 | 67 |
| TA | mp-TiO$_2$/MAPbI$_3$ | λexc = 460 and 750 nm, 2.0 μJ cm$^{-2}$ | Multiexponential fitting | 0.1–0.2 ps | 5000–10000 | 68 |
| TA | mp-TiO$_2$/MAPbI$_3$ | n/a | Multiexponential fitting | 89.6 ps | 11.1 | 62 |

The morphology of the perovskite absorber plays a fundamental role in determining the photovoltaic performance of the device. It directly influences carrier mobility, trap depth and density, charge extraction efficiency, and recombination losses. The microstructure of the active layer, including grain size, degree of crystallinity, uniformity, and packing density, is governed by the synthesis conditions. These include parameters such as temperature, precursor concentration, deposition rate, ambient environment, and the use of passivating additives [69–71].

Strong experimental evidence highlighting the importance of morphology has been provided in reference [72], where the influence of different concentrations of CH$_3$NH$_3$I (MAI) on the structural and photovoltaic properties of MAPbI$_3$ films was systematically studied. These films were fabricated using a two-step deposition method. The study demonstrated that variations in MAI concentration (10 and 40 mg/mL), along with adjustments in annealing conditions and spin-coating parameters, result in significantly different perovskite film morphologies.

At low MAI concentrations, the resulting films exhibited large grain sizes (Figure 2a) but were structurally loose, with poor uniformity and residual PbI$_2$ identified by the characteristic XRD peak at 12.6$^o$. In contrast, a higher concentration of MAI produced dense, smooth, and uniform crystallites (Figure 2b) with a lower content of unreacted precursors and improved substrate coverage (Figure 2c). The surface morphology of MAPbI$_3$ films prepared with low MAI concentrations (Figure 2c) appeared rougher than that of the films fabricated with higher MAI levels (Figure 2d). In addition, the latter exhibited superior substrate coverage. Such morphological features are critically important for achieving high-quality interfaces between the perovskite absorber and the adjacent charge transport layers.

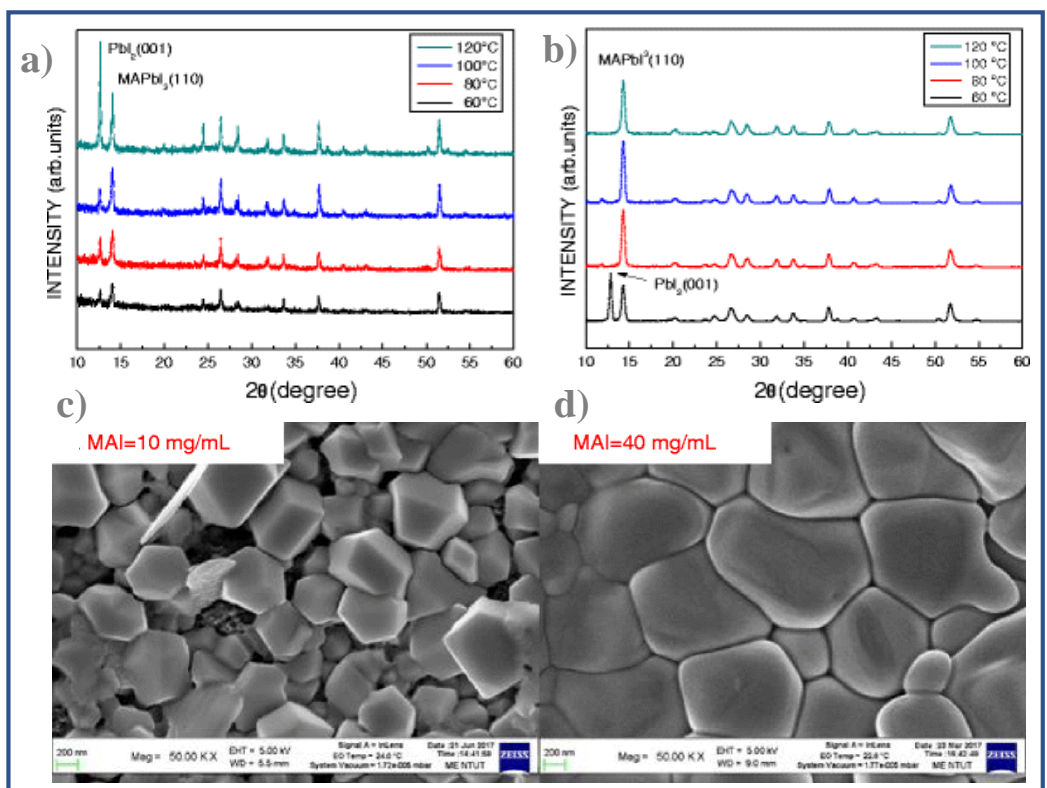

**Figure 2.** XRD patterns of MAPbI$_3$ films synthesized using (a) low and (b) high MAI concentrations; SEM images of MAPbI$_3$ films prepared with (c) low and (d) high-concentration MAI solutions [72].

Photoluminescence (PL) and time-resolved photoluminescence (TRPL) spectroscopy reveal that the denser films prepared with higher MAI concentrations exhibit shorter carrier lifetimes (approximately 14 ns compared to 25 ns in the case of low-concentration films). This observation is attributed to more efficient charge transport and extraction due to a lower trap density and improved contact at the perovskite/TiO$_2$ interface (Figure 3a,b). The redshift observed in PL peak positions also indicates a more complete reaction between PbI$_2$ and MAI, resulting in the formation of a more stoichiometric and defect-free MAPbI$_3$ phase. In this study, the PL peak position was shown to shift from 768 to 773 nm as the MAI concentration increased from 10 to 40 mg/mL (Figure 3a). This redshift is considered to reflect the progress of the chemical conversion of PbI$_2$ into MAPbI$_3$, as previously reported [73]. The interaction of PbI$_2$ films with MAI solution leads to bandgap narrowing in the resulting perovskite layer, with the bandgap shifting toward approximately 1.55 eV. Moreover, the PL intensity of the MAPbI$_3$ film prepared using a higher MAI concentration shows faster decay behavior. The exciton lifetime in such films is relatively short, which helps explain the more rapid exciton dissociation observed.

The interface between TiO$_2$ and the MAPbI$_3$ layer fabricated with a higher MAI concentration appears smooth and continuous, facilitating exciton separation and their efficient extraction to the underlying FTO substrate, as illustrated in Figure 3b. This improved interface quality and film morphology contribute to more rapid electron transport and reduced recombination losses.

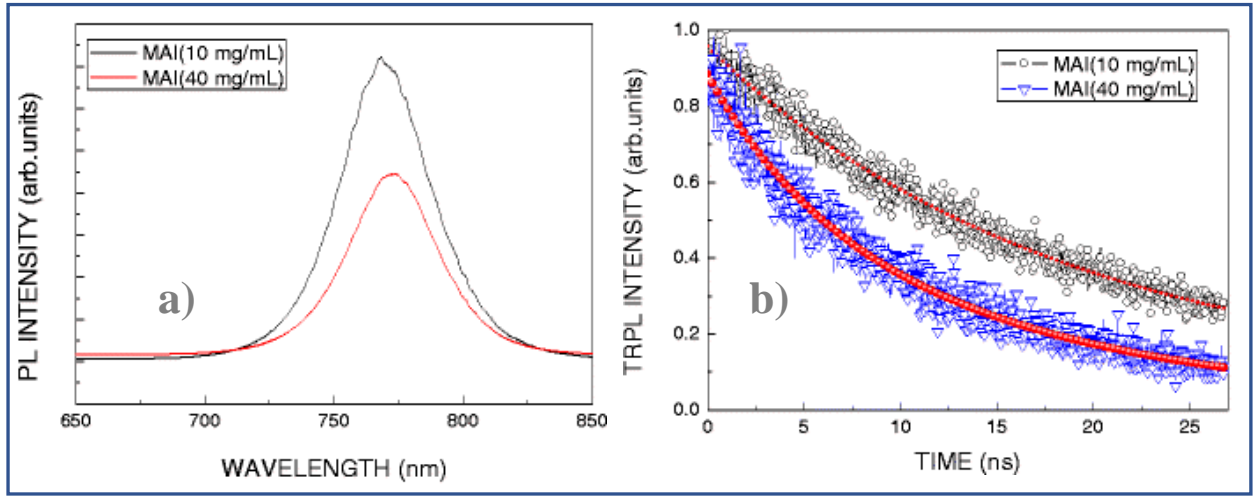

**Figure 3.** Photoluminescence characteristics of MAPbI$_3$ films as a function of MAI concentration: (a) Steady-state photoluminescence (PL) spectra showing emission intensity variations depending on the MAI concentration in the precursor solution. (b) Time-resolved photoluminescence (TRPL) spectra reflecting changes in carrier recombination dynamics for films synthesized with different MAI concentrations [72].

Electrical measurements of J–V characteristics (Figure 4a,b) show that the short-circuit current density (Jsc) increases from approximately 22 mA/cm$^2$ to 24.06 mA/cm$^2$ when the film morphology transitions from loose to compact. This improvement is attributed to a higher fill factor and reduced charge transport resistance, which result from a more ordered crystalline structure. Interestingly, the V$_{oc}$ is slightly higher in devices fabricated with low MAI concentration. This could be explained by the presence of residual PbI$_2$, which may act as a barrier that suppresses interfacial recombination. Nevertheless, the highest efficiency of 17.42% was achieved using a compact morphology with high MAI concentration and an annealing temperature of 120 $^o$C (Figure 4c).

A comparative analysis of power conversion efficiency (PCE) distributions from 50 individual devices (Figure 4c) demonstrates high reproducibility. More than 75% of the devices prepared with high MAI concentration exhibited PCE values above 13% under one-sun illumination. The average PCE values for samples prepared with low and high MAI concentrations were 13.0% and 13.7%, respectively, with similar standard deviations of 1.293% and 1.275%. These findings indicate a stable fabrication process and confirm that the use of high MAI concentration along with controlled annealing temperature leads not only to improvements in Jsc and FF but also to enhanced reproducibility of device performance. It is also noteworthy that devices fabricated using high MAI concentrations consistently exhibit higher perovskite film quality. This results in improved light absorption and, consequently, enhanced photocurrent. In addition, the smooth morphology of these films reduces charge transport resistance, leading to more effective contact between the perovskite layer and the spiro-MeOTAD hole transport layer. This improvement contributes positively to the overall photoresponse of the device [74, 75]. In a related study, the authors of [74] presented a statistical comparison based on histograms of PCE values for 120 devices fabricated with MAPbI$_3$ and MAPbI$_{3-x}$Cl$_x$ films. Their results clearly show that devices based on MAPbI$_{3-x}$Cl$_x$ provided better reproducibility and higher average efficiency of 9.5%, whereas MAPbI$_3$ -based devices displayed a broader and lower distribution of performance (Figure 4d).

Similar improvements in performance have been observed in other passivation strategies. For example, in planar PSCs with surface passivation using PEAI, an efficiency of 23.3% with a

$V_{oc}$ of 1.18 V was achieved. This enhancement is attributed to suppressed grain boundary defects and improved energy level alignment at the perovskite/HTL interface [76]. The use of mixed long-chain ammonium salts, such as tBBAI and PPAI, has also been shown to stabilize the crystal structure and achieve efficiencies as high as 26% [77].

Theoretical studies based on density functional theory (DFT) calculations have further confirmed that the incorporation of cations such as $Cs^+$ or acetylcholine ($Ach^+$) promotes alignment between the conduction and valence bands of the perovskite and charge transport layers. This alignment reduces energetic barriers and enhances carrier extraction efficiency (Figure 4a) [78]. Achieving optimized energy level alignment between the ETL/HTL and the perovskite absorber remains one of the most critical factors for improving $V_{oc}$ in PSCs. Both theoretical predictions and experimental data suggest that a mismatch between the conduction band minimum (CBM) or valence band maximum (VBM) of the transport layers and those of the perovskite can result in internal voltage losses (IVDs), which limit the attainable $V_{oc}$ [79–81]. To reduce such losses, effective passivation materials must possess energy levels that serve as bridges between the perovskite and the transport layers, thereby promoting charge extraction and minimizing recombination.

Furthermore, a study [82] demonstrated that carbonyl and hydrazine groups in benzoyl hydrazine (BH) coordinate effectively with $Pb^{2+}$ ions on the surface of $CsPbI_3$. This interaction eliminates both deep and shallow traps and leads to an upward shift in the CBM, resulting in a $V_{oc}$ increase from 1.17 V to 1.24 V. Similarly, in another investigation [83], acetylcholine ($Ach^+$) was used to compensate for $MA^+$ vacancies. This reduced the density of mid-gap trap states, as confirmed by density of states (DOS) analysis (Figures 4f and 4g). A schematic diagram illustrating the function of an ideal passivator at the ETL/perovskite or HTL/perovskite interface is shown in Figure 4e. The diagram suggests that the passivator modulates either the CBM or the VBM, depending on its location, contributing to an increase in $V_{oc}$. The observed $V_{oc}$ enhancement from 1.12 V to 1.21 V is explained by the elevation of the CBM after defect passivation.

In another study [84], potassium chlorobenzenesulfonate salts were shown to effectively fill oxygen vacancies in $SnO_2$ ETLs. The interaction between the C–Cl bond and unsaturated Sn ions was found to improve energy level alignment, raising $V_{oc}$ from 1.127 V to 1.163 V. Additional work [85] demonstrated that incorporating single-walled carbon nanotubes (SWCNTs) between the $TiO_2$ ETL and the perovskite layer raised the CBM of $TiO_2$ by 0.23 V (Figure 4h) and increased $V_{oc}$ from 0.87 V to 0.93 V. Finally, research [86] confirmed that wide-bandgap oxides such as $Nb_2O_5$ and $Ta_2O_5$ can significantly enhance energy level alignment, resulting in a $V_{oc}$ increase up to 1.16 V.

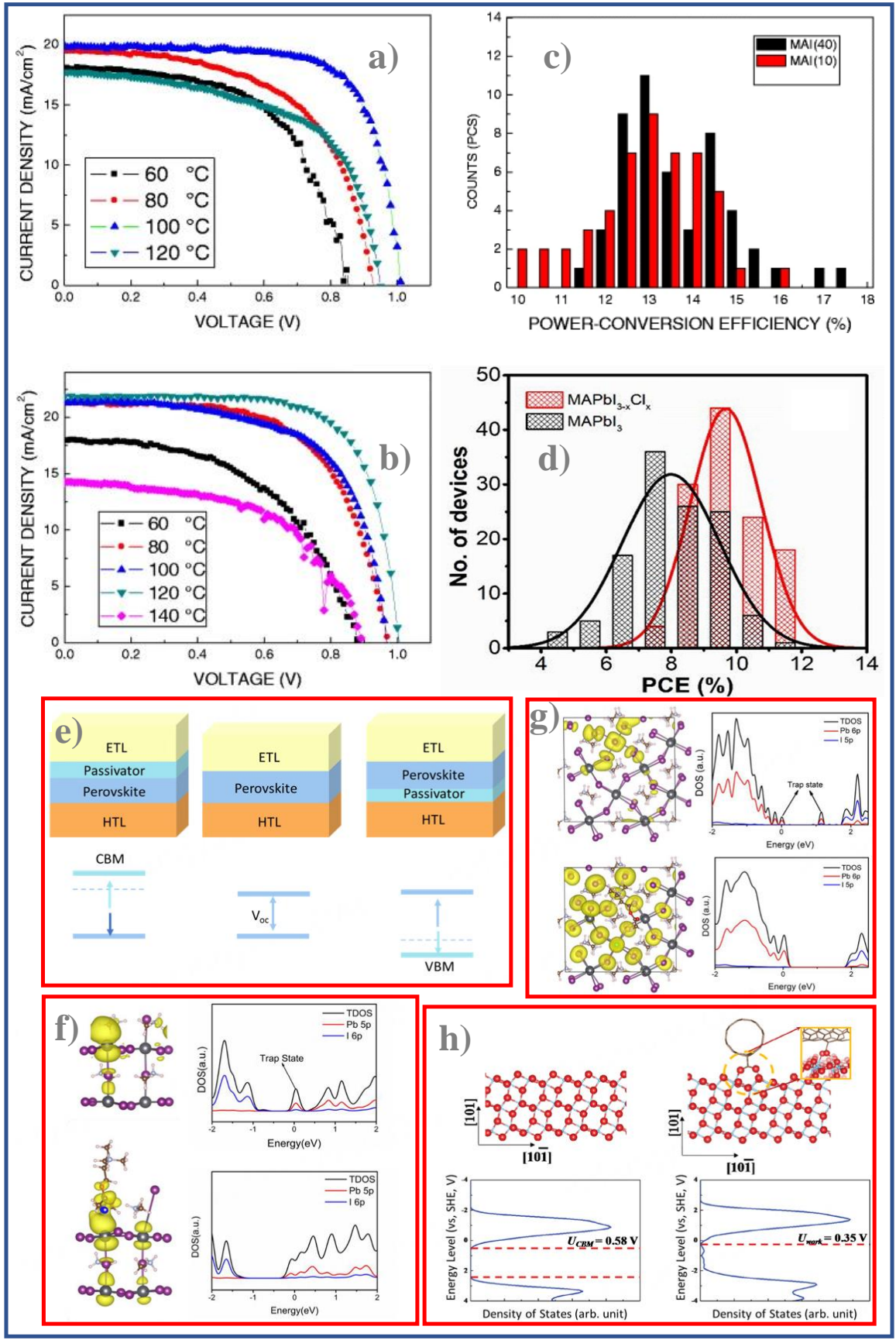

**Figure 4.** J–V curves of perovskite solar cells fabricated using (a) low and (b) high MAI concentration at various annealing temperatures; (c) histogram of PCE values for 50 PSCs fabricated under optimal processing conditions with high MAI concentration [72]; (d) PCE histogram for 120 devices with an active area of 0.18 cm$^2$ [74]; (e)

schematic diagram showing the role of an ideal passivator at the ETL/perovskite or HTL/perovskite interface. The passivator adjusts the position of the conduction band minimum (CBM) or valence band maximum (VBM), contributing to an increase in $V_{oc}$. (f) $Cs^+$ passivator filling $MA^+$ vacancies results in a downward shift of the VBM and an increase in $V_{oc}$; (g) use of $Ach^+$ to passivate $MA^+$ vacancies eliminates deep traps, as confirmed by density of states (DOS) analysis [83]; (h) $Ach^+$ also helps eliminate shallow traps within the forbidden energy range, as evidenced by DOS analysis [85].

At the perovskite/HTL interface, 3-(aminomethyl) pyridine (3-APy) was employed in [87] to reduce surface roughness and potential fluctuations. DFT calculations showed that iodine vacancies act as n-type dopants, shifting the VBM from 0.80 to 1.51 eV and increasing $V_{oc}$ up to 1.19 V. In another study [88], Cs-doped MXene $Ti_3C_2T_x$ passivators restored the VBM to −5.90 eV due to strong adsorption of $Cs^+$ into $MA^+$ vacancies (with an adsorption energy of −5.13 eV), which led to a $V_{oc}$ increase from 1.05 to 1.10 V. The work in [89] employed BMBC to enhance coordination with $Pb^{2+}$, improving energy level alignment and raising $V_{oc}$ to 1.249 V. Additionally, trifluoroacetamidine was used in [90] to interact with the perovskite via hydrogen bonding between fluorine-containing groups and the crystal surface, thereby reducing grain boundary recombination and increasing $V_{oc}$ from 1.10 to 1.16 V.

Interface engineering also includes the use of buffer oxide layers. For example, the introduction of $Y_2O_3$ between the perovskite and the electrode was found to increase carrier lifetime and suppress reverse current, as confirmed by TRPL measurements [91]. Such barrier layers help reduce interfacial trap densities and improve operational stability under ambient conditions. Combining morphological control with chemical passivation enables not only high PCE values but also excellent device reproducibility. In the study by [72], carried out on a set of 50 devices, the standard deviation of PCE was only about 1.3%, and more than 75% of the devices exhibited efficiencies above 13%, confirming the reliability of the process under optimized morphological conditions.

### 3. Strategies for Enhancing Charge Transport

Controlling charge transport is a key area in the development of perovskite solar cells (PSCs), as the energy conversion efficiency directly depends on how effectively photogenerated carriers can be extracted from the active layer and delivered to the electrodes. The main challenge is not only to ensure favorable pathways for both electron and hole transport but also to minimize recombination losses that occur during charge migration. Achieving efficient transport requires a comprehensive approach involving the optimization of electron and hole transport layers (ETL and HTL), defect management at interfaces, morphological tuning of the active layer, and architectural design at the device level. Among the most extensively studied and promising ETL materials is tin dioxide ($SnO_2$), which has increasingly replaced the traditionally used titanium dioxide ($TiO_2$). $TiO_2$ suffers from relatively low electron mobility and is prone to inducing hysteresis in current–voltage measurements. $SnO_2$, in contrast, offers higher electron mobility and more favorable band alignment with perovskite materials such as $MAPbI_3$ and $FAPbI_3$, making it an attractive choice for modern PSC architectures [92, 93].

In the work by Yoo et al. [94], an improved synthesis method for $SnO_2$-based ETLs was proposed using chemical bath deposition (CBD). The authors showed that at stage A-ii (pH ≈ 1.5), a uniform, dense, and nearly defect-free film of approximately 50 nm thickness was formed. This film provided full coverage of the substrate and minimized interfacial losses. Such a layer promotes efficient electron extraction and suppresses hole back-transfer, which is crucial for enhancing PSC performance. With the optimized $SnO_2$ ETL, the devices achieved a certified record efficiency of 25.2%, with a fill factor exceeding 84% and an open-circuit voltage of 1.225 V.

The morphological and structural properties of SnO$_2$ layers at various deposition stages, including TEM and XRD data, are shown in Figures 5a–e. High-resolution electron microscopy (Figures 6a,b) confirmed that the SnO$_2$ film uniformly covered the textured surface of the FTO substrate, including recessed regions, thereby ensuring effective electron extraction and hole blocking [94].

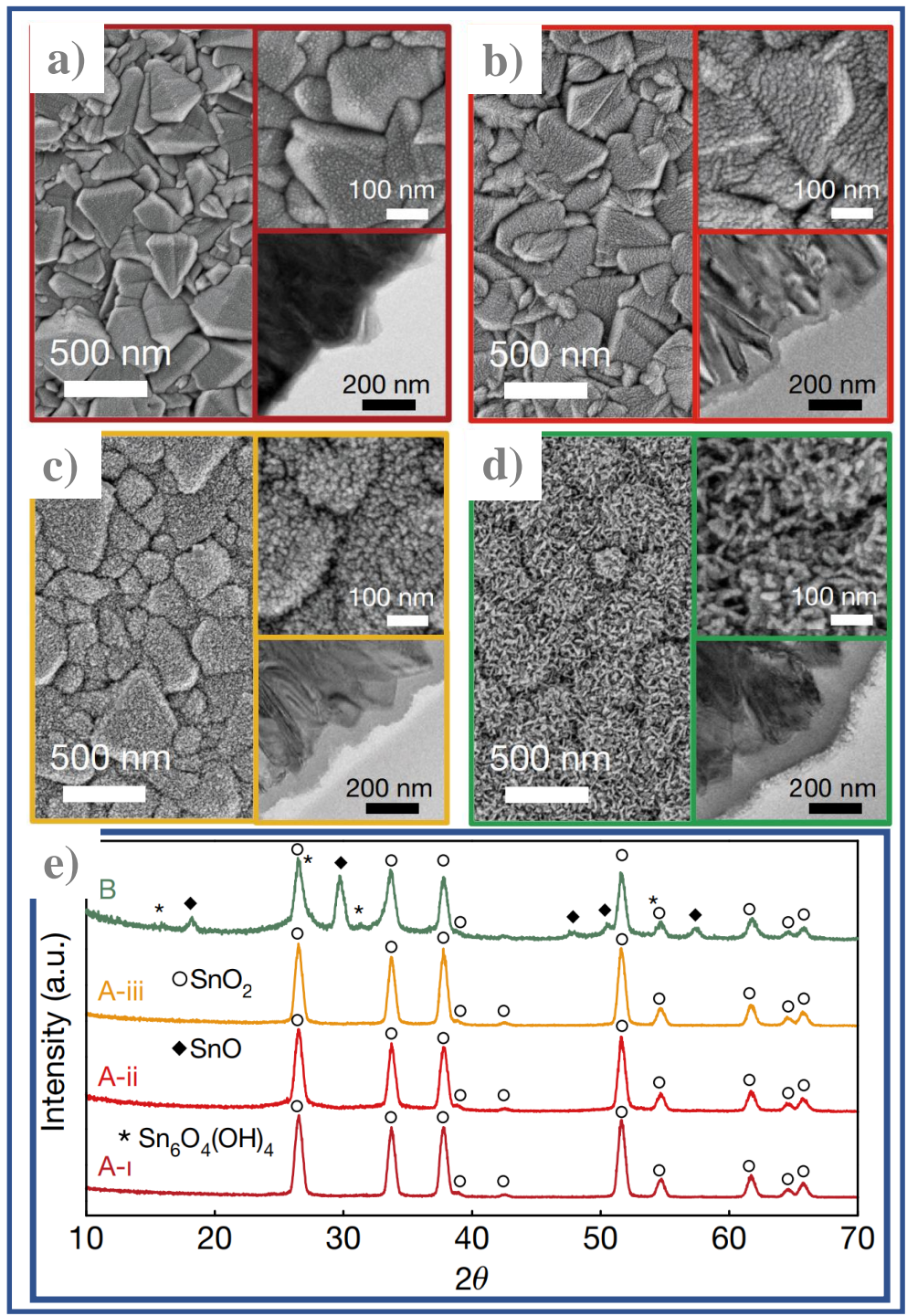

**Fig. 5.** Synthesis and characterization of tin oxide film on FTO. (a−d): Top-view SEM images synthesized for increasing time, leading to an increased pH: a), stage A-i; b), stage A-ii; c), stage A-iii; and d), stage B. The top-right insets show a zoomed-in SEM and the bottom-right insets show the corresponding cross-sectional TEMs. e) X-ray diffraction patterns of the SnO$_2$ layer at different reaction stages. a.u., arbitrary units [94].

Based on their experimental observations, the authors classified the reaction into four distinct stages (Figure 6c). In the initial phases (stages A-i and A-ii, pH = 1–1.5), the main product of chemical deposition was identified as $SnO_2$. As the process transitioned to stage A-iii, a sharp decrease in dissolved oxygen concentration led to incomplete oxidation of $Sn^{2+}$ to $Sn^{4+}$. This condition favored the formation of non-stoichiometric tin oxide, $SnO_{2-x}$, where $0 < x < 2$, as also discussed in [95]. During this stage, the deposition solution became visibly cloudy, which indicated the formation of insoluble amorphous tin oxide species in the bath [95].

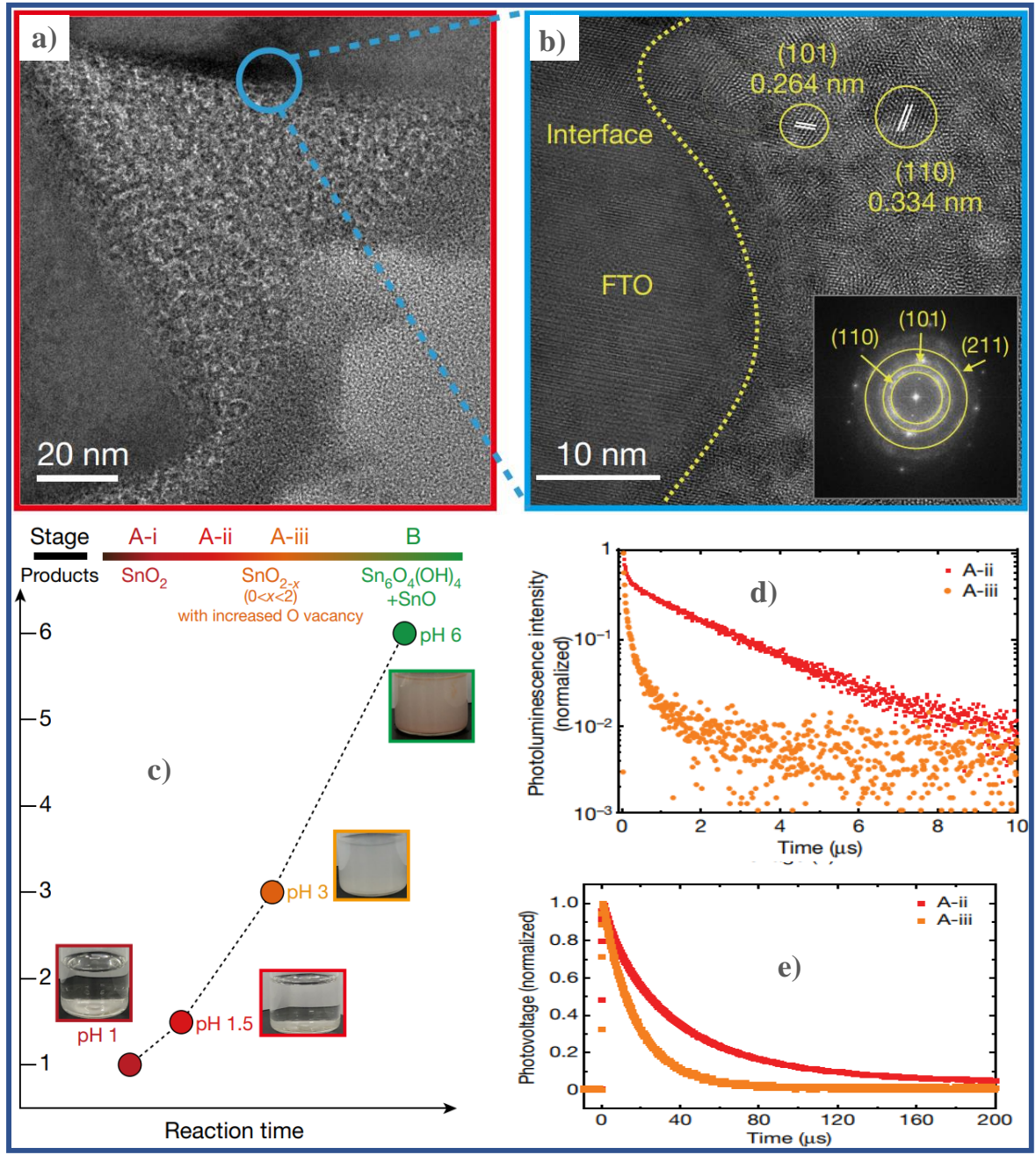

**Fig. 6.** Synthesis stages and optoelectronic characterization of $SnO_2$ films on FTO substrates. (a, b) High-resolution TEM images of $SnO_2$ films prepared up to stage A-ii; inset in (b) shows FFT pattern indicating crystallographic planes. (c) Schematic of reaction progress showing corresponding pH changes, products, and solution appearance at each stage (A-i to B). (d) Time-resolved photoluminescence (TRPL) spectra of perovskite films deposited on $SnO_2$ from stages A-ii (red) and A-iii (orange); $\tau_{avg}$ = 984 ns for A-ii and 81 ns for A-iii, indicating suppressed non-radiative recombination at the A-ii interface. (e) Transient photovoltage decay curves of PSCs based on A-ii and A-iii SnO$_2$ layers; decay times of 37.8 μs and 17.9 μs, respectively, further confirm lower recombination rates for A-ii [94].

An important factor contributing to the enhancement of device performance is the control of defects within the charge transport layers and at their interfaces with the perovskite absorber. The study [94] demonstrated that increasing the pH value above 3, corresponding to stage A-iii of the deposition process, leads to the formation of secondary phases such as Sn$_6$O$_4$(OH)$_4$ and SnO. It also promotes the formation of oxygen vacancies, which act as trap states and intensify non-radiative recombination. This behavior is supported by the reduction in carrier lifetime measured using TRPL (Figure 6d). Similarly, transient photovoltage measurements (Figure 6e) confirmed faster voltage decay in devices with defective ETLs, indicating enhanced recombination processes. Comparable results have been reported in other studies. For instance, the use of self-assembled monolayers (SAMs) at the ETL/perovskite interface was shown to increase $V_{oc}$ by 75 mV and improve the fill factor by 4% [96].

Another effective approach involves modifying the perovskite active layer by incorporating dopants. As noted earlier in Section 3, Yoo et al. [94] also investigated compositional tuning of FAPbI$_3$ films through the introduction of small amounts of MAPbBr$_3$. Their experiments showed that adding just 0.8 mol% of MAPbBr$_3$ significantly increased grain size, as observed in SEM images (Figure 7a). This additive also stabilized the desired α-phase of FAPbI$_3$ and suppressed the formation of the non-photoactive δ-phase. The resulting morphological improvement was accompanied by an increase in carrier lifetime to 3.6 μs (Figure 7b), along with an effective charge carrier mobility of up to 31.2 cm$^2$/V·s, as measured by optical pump–terahertz probe spectroscopy (Figure 7c). These values are comparable to the best results reported for 2D interface passivation and targeted perovskite doping strategies [97, 98].

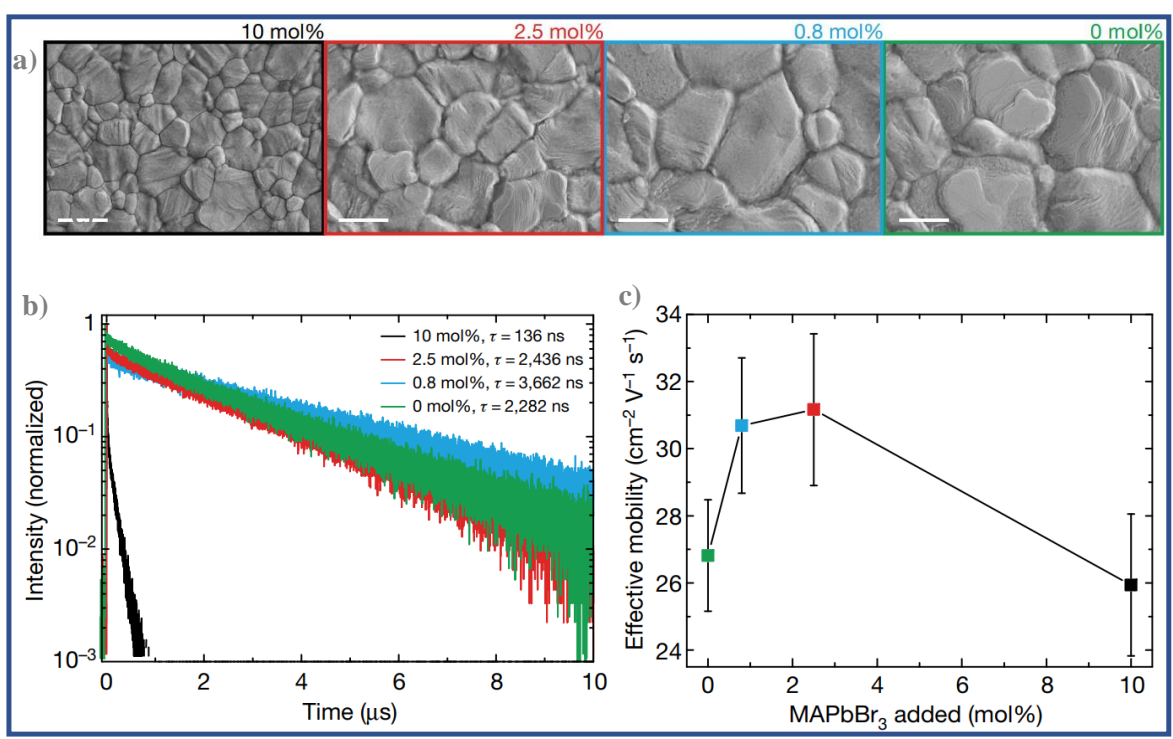

**Fig. 7.** Characterization of perovskite thin films varying the mole per cent of MAPbBr$_3$ added. a), SEM images of perovskite thin films with decreasing amounts of added MAPbBr$_3$ (from 10% to 0%). The scale bars are 1 μm. b), Time-resolved photoluminescence traces for perovskite thin films with decreasing amounts of added MAPbBr$_3$.

Carrier lifetimes τ are obtained from a mono-exponential fit. c), Effective mobility for perovskite thin films determined from optical pump−terahertz probe measurements (error bars show one standard deviation) [94].

Architectural strategies such as nanostructuring of the electron transport layer (ETL) and the creation of gradient energy profiles between functional layers are becoming increasingly common. The use of buffer layers like PEAI or $C_{60}$ at the ETL/perovskite interface has been shown to reduce surface trap densities and improve carrier extraction efficiency [99]. In a related study [100], these interfacial buffer layers enhanced energy level alignment and suppressed leakage currents, resulting in an increase in $V_{oc}$ by more than 100 mV.

In addition, the development of cascade-type 3D/2D hybrid perovskite structures not only stabilizes the crystal phase but also effectively passivates grain boundary defects. This contributes to improved device resistance to both moisture and thermal stress [101]. A comprehensive evaluation of these approaches and their quantitative impact on device parameters is presented in summary form in Table 2. The table consolidates the effects of various strategies on charge carrier lifetime, open-circuit voltage, fill factor (FF), and power conversion efficiency (PCE) in perovskite solar cells (PSCs) [94, 97–101]. For comparative interpretation, the table includes the values of ΔFF and ΔPCE, which indicate the improvement in these key performance parameters relative to the untreated control samples.

Table 2. Effects of Charge Transport Optimization Strategies on PSC Performance Parameters

| **Strategy** | **Effect on τ (TRPL)** | **ΔV$_{oc}$** | **ΔFF** | **ΔPCE** | **Ref.** |
|---|---|---|---|---|---|
| SnO$_2$, stage A-ii (CBD, pH ≈ 1.5) | ↑ up to 984 ns | +90 mV | +5 % | +3–5 % | [94] |
| SnO$_2$, stage A-iii (defective) | ↓ down to 81 ns | −120 mV | −6 % | −4–5 % | [94] |
| Addition of 0.8 mol% MAPbBr$_3$ | ↑ up to 3.6 μs | +30 mV | +2 % | +2–3 % | [94] |
| SAM interface modification (C$_{60}$-SAM) | - | +75 mV | +4 % | - | [96] |
| PEAI buffer layer between ETL and perovskite | - | +100 mV | +4–5 % | +2–4 % | [99] |
| 3D/2D hybrid structures (bilayer phases) | ↑ up to 1.5 μs | +50 mV | +3 % | +2–3 % | [97], [101] |

An analysis of the data presented in Table 2 clearly shows that each modern strategy for optimizing charge transport has a distinct effect on the photovoltaic performance of PSCs. For example, optimized deposition of SnO$_2$ improves both carrier lifetime and output voltage. In contrast, the use of SAM modifiers or PEAI buffer layers primarily affects energy level alignment and enhances the fill factor. The addition of MAPbBr$_3$, on the other hand, promotes grain growth and suppresses bulk defects, leading to simultaneous improvements across all key parameters.

It is important to emphasize that no single strategy is universally effective. Only by combining multiple approaches can a synergistic effect be achieved. For instance, integrating high-quality SnO$_2$ with interfacial passivation using PEAI and morphological control via Br⁻ additives has resulted in devices with record values of $V_{oc}$, FF, and long-term stability. Enhancing charge transport in PSCs requires an integrated approach that combines precise ETL engineering, defect suppression, energy level alignment, optimization of film morphology, and the use of buffer and passivation layers. This combination allows devices to approach their theoretical limits in terms of $V_{oc}$, FF, and PCE, while also significantly improving operational stability under real-world conditions.

## 4. Suppression of Recombination Losses in PSC

Recombination losses remain one of the main obstacles to maximizing the efficiency of perovskite solar cells. The primary recombination mechanisms include bulk recombination within the active layer, interfacial recombination at the perovskite/ETL and perovskite/HTL interfaces, and trap-assisted recombination due to crystal lattice defects and grain boundaries. These losses lead to reductions in key device parameters such as open-circuit voltage, fill factor (FF), and power conversion efficiency (PCE). To address these limitations, a range of strategies have been developed, including passivation techniques, interface engineering, morphological control, and structural modifications, all aimed at reducing the probability of recombination.

One of the most significant achievements in reducing interfacial recombination has been the use of $SnO_2$ layers fabricated via chemical bath deposition (CBD) with controlled pH conditions. As discussed in Section 3, the study by Yoo et al. [94] showed that at pH ≈ 1.5 (stage A-ii), a high-quality, dense, and defect-tolerant $SnO_2$ layer was formed. This layer provided uniform substrate coverage and reliable contact with the perovskite. Time-resolved photoluminescence (TRPL) measurements revealed a sharp increase in carrier lifetime ($\tau\_PL$) up to 984 ns, compared to just 81 ns for the $SnO_2$ film deposited at pH ≈ 3 (stage A-iii) (Figures 6c and 6d). In addition, transient photovoltage (TPV) measurements confirmed that voltage relaxation in the A-ii structure was almost twice as slow as in the A-iii structure, further indicating reduced recombination (Figure 6d). These results highlight the importance of film purity and phase stability in the ETL for minimizing interfacial losses.

Another critical focus area is passivation at the upper interface between the perovskite and the hole transport layer (HTL). Multiple studies have shown that incorporating self-assembled monolayers (SAMs) or organic buffer layers such as PEAI significantly reduces surface defect density and improves energy level alignment. The use of a SAM layer between $SnO_2$ and the perovskite has been reported to increase $V_{oc}$ by 75 mV and improve FF by 4% [102]. Similarly, Tan et al. [99] achieved a $V_{oc}$ increase of approximately 100 mV using PEAI as a buffer layer. This enhancement was attributed to suppression of $Pb^o$-related defects and better band alignment (Figures 8a–c). In addition, recent work has shown that replacing conventional $TiO_2$ with modified $TiO_2$-Cl in planar PSCs improves long-term device stability due to enhanced interfacial bonding and reduced charge accumulation at the ESL/perovskite boundary [99, 103–105]. For example, MAFA-based devices using $TiO_2$-Cl retained up to 95% of their initial PCE after 60 days of dark storage, whereas devices based on unmodified $TiO_2$ retained only 38% (Figure 8a). Furthermore, high-efficiency CsMAFA devices (PCE > 21%) on $TiO_2$-Cl maintained 96% of their initial performance even after 90 days (approximately 2000 hours). Under continuous operation at the maximum power point (MPP), these cells showed strong stability, with PCE dropping to only 90% after 500 hours of 1-sun illumination under UV-filtered conditions (Figure 8b). This effect is partly attributed to early-stage healing of light-induced defects, followed by slower degradation at the perovskite/spiro-OMeTAD interface. After dark storage, the efficiency partially recovered to 97% of its initial value (Figure 8c), confirming the reversibility of some degradation processes.

A particularly notable example of effective molecular passivation was reported by Osman et al. [106], who used n-octylammonium iodide (OAI) to treat the perovskite surface. The authors found that OAI reduced trap density by a factor of fifty and significantly increased the hole capture time at the perovskite/HTL interface (Figure 8d). They also observed a marked improvement in film morphology upon adding 2 mg/mL of OAI, with average surface roughness decreasing from approximately 42.6 nm to 32.8 nm according to AFM measurements (Table 3). This led to tighter interfacial contact between the perovskite and Spiro-OMeTAD, reduced interfacial resistance, and thus a lower likelihood of recombination.

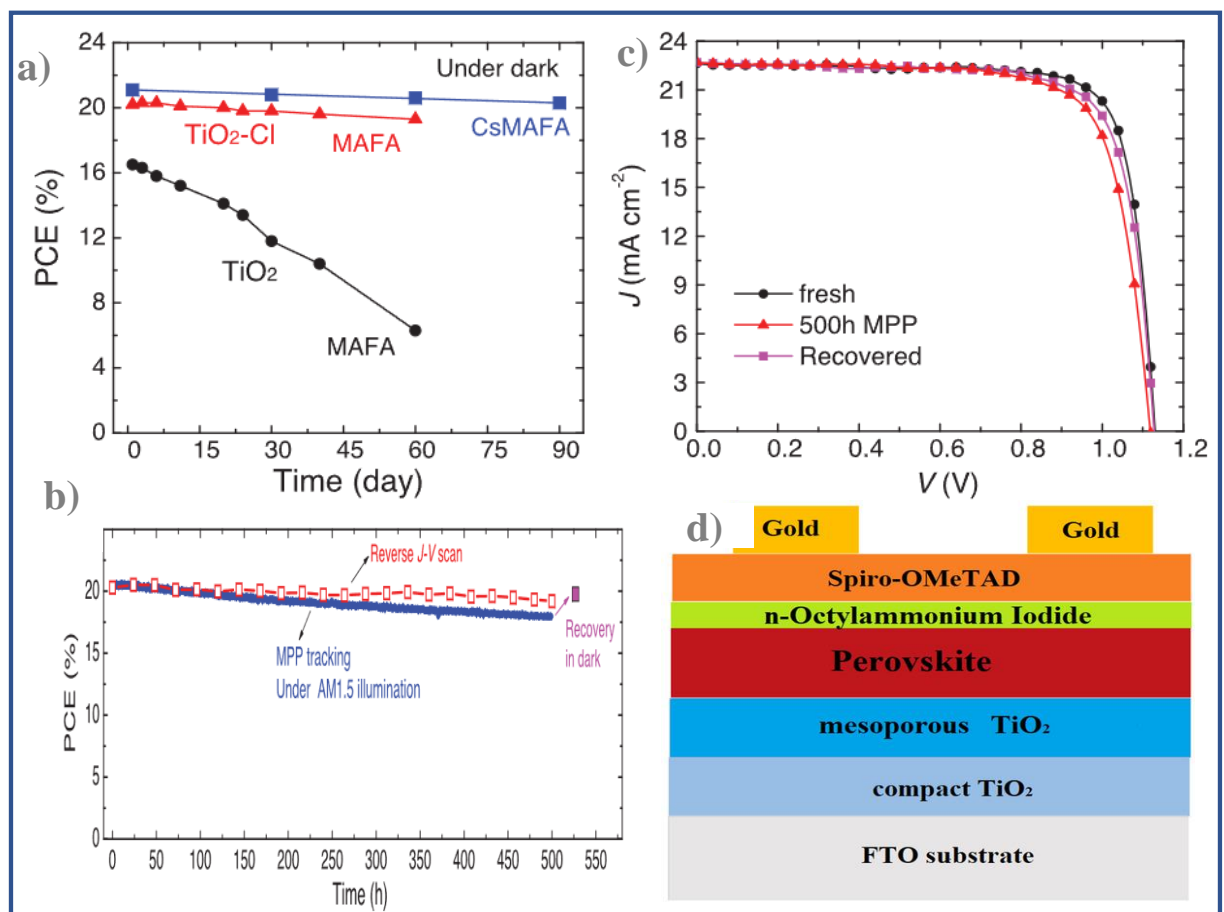

**Figure 8.** Long-term stability of perovskite solar cells (PSCs) fabricated with $TiO_2$ and $TiO_2$-Cl. (a) Shelf-life performance of unencapsulated PSCs based on $TiO_2$ and $TiO_2$-Cl under dark storage. Devices were stored in a dry box (humidity <30%) in the dark and measured periodically in a nitrogen atmosphere. The reported power conversion efficiency (PCE) values were obtained from reverse-scan J–V measurements. (b) Maximum power point (MPP) tracking over 500 hours for a high-performance unencapsulated CsMAFA-based device with $TiO_2$-Cl, under nitrogen atmosphere and continuous simulated sunlight (100 mW/cm$^2$) using a 420 nm cutoff UV filter. PCE values extracted from reverse J–V scans are also shown (square symbols); the device retains 95% of its initial efficiency. (c) J–V curves of the CsMAFA device recorded at different stages of the test [99]. (d) Schematic device architecture of the PSC passivated with n-octylammonium iodide [106].

Wu et al. [107] demonstrated that treating the perovskite/$C_{60}$ interface with $PDMAI_2$ effectively reduces trap density by half, which enhances both the photoluminescence quantum efficiency (PLQE) and the overall PCE, reaching values up to 25.5% (Figures 9a–c). The proposed device architecture with interfacial passivation is illustrated in Figure 9a. A direct comparison between passivated and unpassivated devices showed that the introduction of OAI led to a significant increase in $V_{oc}$. This improvement was attributed to suppressed surface recombination through defect states (Figure 9b). An increase in PLQE from 4.9% to 13.1% further confirmed the effectiveness of interfacial defect passivation (Figure 9c). The authors explained this effect by suggesting that OAI molecules passivate both $FA^+$ and $I^-$ vacancies at the interface. This passivation occurs without the formation of a two-dimensional phase, as confirmed by low-angle X-ray diffraction analysis. This study underscores the importance of fine-tuning interface properties using organic ammonium species in order to suppress trap states and enhance the operational stability of perovskite solar cells.

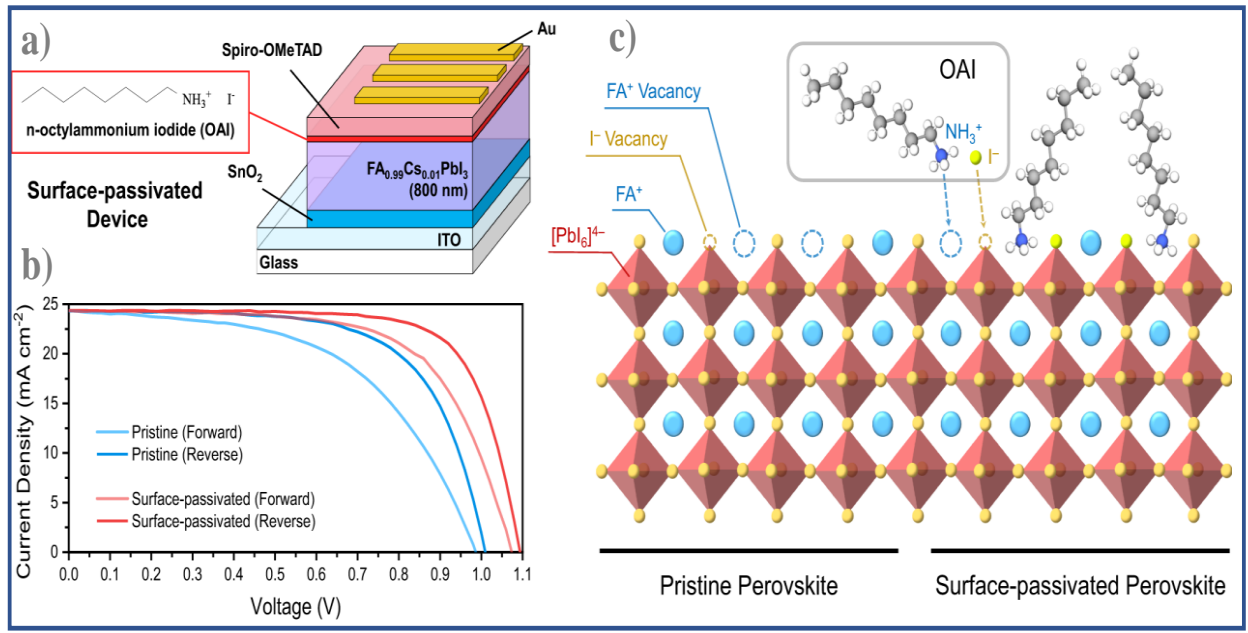

**Figure 9. a)** device architecture of the surface-passivated device. The pristine device has a similar architecture but without the OAI layer. **b)** Current density versus voltage of the pristine and surface-passivated devices measured during forward and reverse voltage sweeps at a scan rate of 40 mV s$^{-1}$. **c)** Configuration of the surfaces of pristine and surface-passivated perovskites. The OA chains mount to the perovskite surface via their $NH_3^+$ groups, which also passivate the $FA^+$ vacancies simultaneously, while the $I^-$ vacancies can be passivated by the $I^-$ in OAI [107].

**Table 3.** $R_a$ and RMS values determined by AFM for the perovskite $(MACl)_{0.33}FA_{0.99}MA_{0.01}Pb(I_{0.99}Br_{0.01})_3$ film with n-octylammonium iodide concentrations [106].

| Perovskite Films | $R_a$ (nm) | RMS (nm) |
|---|---|---|
| $(MACl)_{0.33}FA_{0.99}MA_{0.01}Pb(I_{0.99}Br_{0.01})_3$ | 42.6 | 53.5 |
| $(MACl)_{0.33}FA_{0.99}MA_{0.01}Pb(I_{0.99}Br_{0.01})_3$/2OAI | 32.8 | 40.6 |
| $(MACl)_{0.33}FA_{0.99}MA_{0.01}Pb(I_{0.99}Br_{0.01})_3$/4OAI | 34 | 42.6 |
| $(MACl)_{0.33}FA_{0.99}MA_{0.01}Pb(I_{0.99}Br_{0.01})_3$/6OAI | 36.6 | 51.2 |

Time-resolved photoluminescence (TRPL) measurements revealed that the carrier lifetime increased from 1561 ns to 2463 ns following passivation (Table 4), which is consistent with the observed enhancement in photoluminescence intensity (Figure 10a). The J–V characteristics (Figure 10b) demonstrated an increase in $V_{oc}$ from 1.02 V to 1.06 V, a rise in fill factor from 75% to 79%, and an overall power conversion efficiency reaching 20.2%. These results confirm that even simple molecular passivation can have a significant impact on PSC performance by effectively suppressing both interfacial and trap-assisted recombination.

**Table 4.** TRPL lifetime determined by mono-exponential fitting control and target films deposited on glass [106].

|       | $t_1$ (ns)        |
|-------|-------------------|
| 0 OAI | 1561.20 ± 42.63   |
| 2 OAI | 2462.52 ± 26.67   |

Controlling the morphology of the perovskite layer also plays a critical role in device performance. As shown in Figure 7b, the addition of 0.8 mol% MAPbBr$_3$ to FAPbI$_3$, aimed at suppressing the δ-phase and increasing grain size, resulted in a substantial improvement in carrier lifetime, reaching 3.6 μs, and charge mobility, which increased to 31.2 cm$^2$/V·s [94, 108]. A comparable effect can be achieved by incorporating 3D/2D hybrid structures, which facilitate surface state passivation and enhance device stability. Liu et al. [97] demonstrated that the introduction of 2D components such as PEA$_2$PbI$_4$ led to a $V_{oc}$ of approximately 1.19 V, a carrier lifetime (τ_PL) of 1.6 μs, and a power conversion efficiency of up to 22.3%.

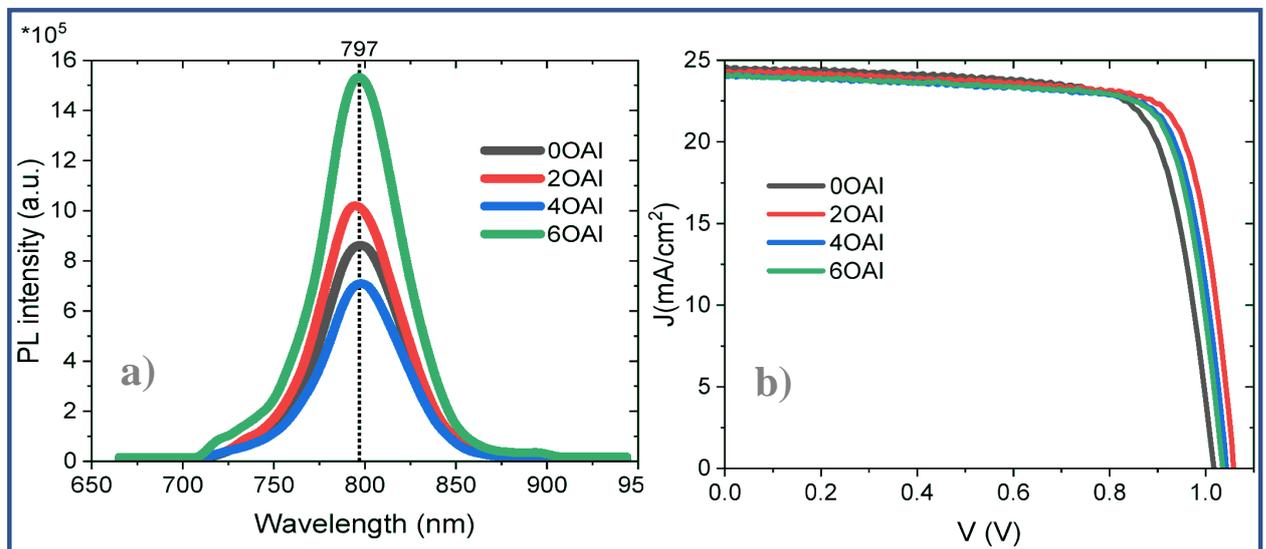

**Figure 10.** (a) PL spectra of the perovskite (MACl)$_{0.33}$FA$_{0.99}$ MA$_{0.01}$Pb(I$_{0.99}$Br$_{0.01}$)$_3$ films with n-octylammonium iodide concentrations and (b) Current density−voltage (J–V) curves of the PSCs with n-octylammonium iodide concentrations [106].

At the fundamental level of recombination mechanisms, the study by Stolterfoht et al. [109] demonstrated that the primary losses in $V_{oc}$ are associated with interfaces rather than the bulk of the material. This conclusion was supported by quasi-Fermi level splitting measurements. As a result, optimizing the interfaces between the perovskite absorber and the charge transport layers remains one of the most important objectives. Strategies aimed at controlling trap-assisted recombination also play a vital role. For example, surface passivation using APTMS has been shown to reduce trap density and lower the surface recombination velocity to 125 cm/s. This was accompanied by an increase in $V_{oc}$ from 1.03 V to 1.09 V [110, 111]. The use of guanidinium as a passivating agent, according to DFT calculations, helps stabilize surface states and improves the fill factor from 20.3 percent to 22.5 percent [112]. In addition, CF$_3$-PEA has been shown to increase the fill factor from 80.7 percent to 82.8 percent [78, 113]. This combined theoretical and experimental approach highlights the potential of using machine learning and DFT-based tools for the optimization of materials used in PSCs.

Recent advances in recombination suppression confirm that only an integrated strategy can lead to record efficiencies. This includes coordinated control over phase stability, film morphology, interfacial quality, and synthesis conditions. Such a multifaceted engineering approach, beginning with solution chemistry and extending to quantum-level modeling, defines the pathway toward next-generation perovskite solar cells that combine high efficiency, long-term stability, and minimal recombination losses.

## 5. Optimizing Carrier Lifetime and Mobility

Optimizing the charge carrier lifetime ($\tau$) and mobility ($\mu$) is considered one of the most essential tasks in the field of perovskite solar cells. Improvements in these parameters contribute directly to better $V_{oc}$, higher fill factor (FF), and increased overall power conversion efficiency (PCE). Recent progress has been made through the combined use of advanced diagnostics, selective chemical strategies, and innovative device architectures, forming the technological foundation that supports the move toward commercialization.

A number of time-resolved diagnostic methods, including TRPL, TAS, SCLC, OCVD, ECPL, and TRMCD, have contributed significantly to our understanding of recombination and transport mechanisms. In particular, Bojtor et al. [114] applied the time-resolved microwave conductivity decay (TRMCD) method to separate carrier lifetime contributions into radiative, Shockley–Read–Hall, and Auger components. Their results showed that even under high illumination, the dominant limitation remained SRH recombination, as illustrated in Figure 11. The study further demonstrated that logarithmic binning of the recorded signal allows effective analysis of carrier decay over a broad timescale, from hundreds of nanoseconds to several milliseconds. This approach combines excellent temporal resolution with improved signal-to-noise ratio but requires careful selection of the binning interval. If the intervals are too small, fast recombination components may be averaged out. If they are too large, the slower decay phases may become distorted or smoothed.

Despite these considerations, logarithmic binning (Figure 11d) enables consistent dynamic sensitivity across the full time domain, capturing data with high accuracy from the earliest nanosecond-scale processes to the long-tail millisecond region. The voltage decay curve averaged over logarithmic bins (Figure 11a), together with the corresponding carrier density dynamics (Figures 11b and 11c), illustrates the method's effectiveness in resolving complex recombination behavior.

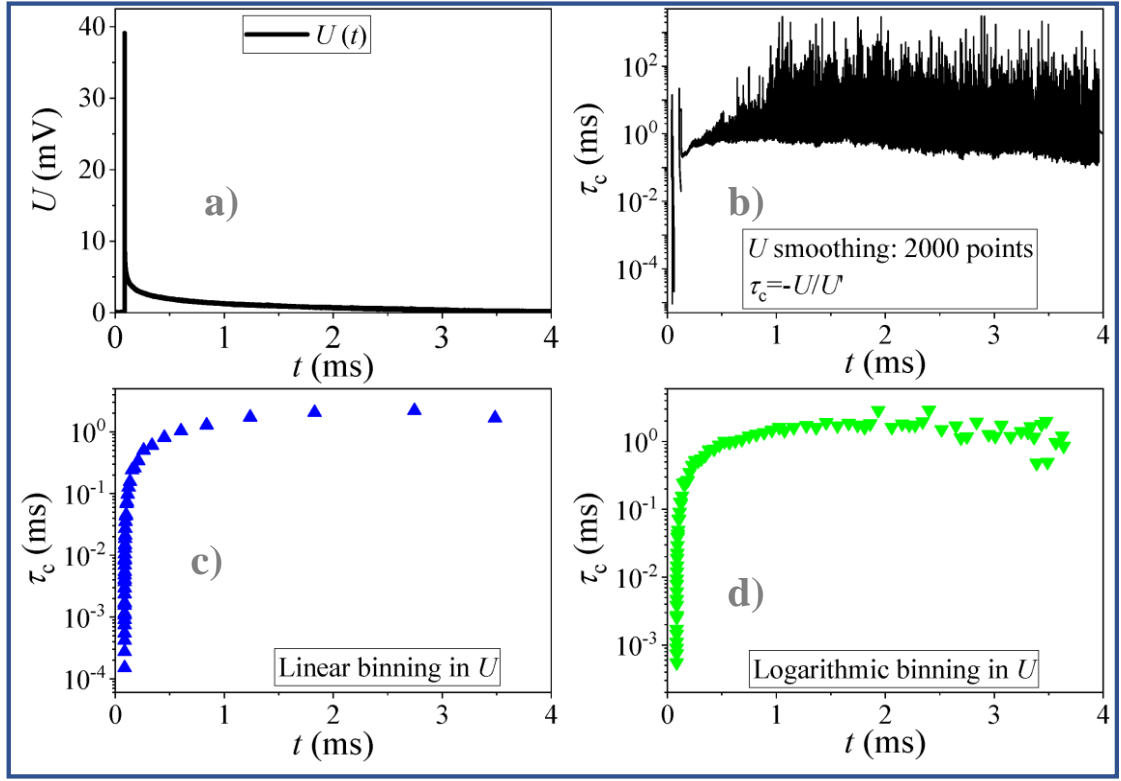

**Figure 11.** (a) Raw voltage signal in the time domain with binning-averaged values overlaid. (b) Charge carrier recombination dynamics calculated after signal smoothing and numerical differentiation. (c) Recombination dynamics derived from data binned into linearly spaced intervals. (d) Recombination dynamics obtained using logarithmically spaced bins [114].

Rodriguez-Perez et al. [115] applied open-circuit voltage decay (OCVD) analysis to quantitatively assess ionic mobility in triple-cation perovskites with the composition $FA_xMA_{1-x}Cs_xPb(I_{0.8}Br_{0.2})_3$. They reported ionic mobilities on the order of $10^{-10}$ to $10^{-12}$ cm$^2$/V·s and demonstrated that increasing the proportion of FA$^+$ ions and optimizing the crystallization process leads to an extended carrier lifetime ($\tau$) and improved device stability, as summarized in Table 5.

Table 5. Combined summary of photovoltaic performance and equivalent circuit parameters for solar cells with various concentrations of $Cs_{0.05}FA_{1-x}MA_xPb(I_{1-x}Br_x)_3$ [115].

| | **Photovoltaic parameters of $Cs_{0.05}FA_{1-x}MA_xPb(I_{1-x}Br_x)_3$ solar cells** | | | | | | | |
|---|---|---|---|---|---|---|---|---|
| FAI | Rapid increase | | Slow increase | | Fast decay | | Slow desay | |
| M | $\Delta V_{R1}$ (V) | $\tau_{r1}$ (ms) | $\Delta V_{R2}$ (V) | $\tau_{r2}$ (s) | $\Delta V_{D1}$ (V) | $\tau_{d1}$ (ms) | $\Delta V_{D2}$ (V) | $\tau_{d2}$ (s) |
| 1.135 | 1.07 | 90 | 0.36 | 9.42 | 0.445 | 98 | 0.39 | 83.53 |
| 1.03 | 0.968 | 140 | 0.03 | 9.43 | 1.009 | 173 | 0.076 | 82.67 |
| 0.924 | 0.817 | 90 | 0.08 | 9.65 | 0.714 | 86 | 0.296 | 83.86 |

| | **Equivalent circuit parameters ($Cs_{0.05}FA_{1-x}MA_xPb(I_{1-x}Br_x)_3$)** | | | | | |
|---|---|---|---|---|---|---|
| FAI | Rrec | Rt | R, bulk | Ri, bulk | Ri, acc | Racc |
| M | Ωcm² | Ω cm² | F cm² | Ω cm² | Ω cm² | F cm² |
| 1.135 M | 2.54 ×10$^{12}$ | 0.50 ×10$^9$ | 3.99 ×10$^{-2}$ | 25 ×10$^9$ | 25.0×10$^9$ | 1.56 ×10$^{-3}$ |
| 1.03 M | 8.8 × 10$^8$ | 1.00 ×10$^3$ | 7.18 ×10$^{-1}$ | 50 ×10$^3$ | 50.0×10$^3$ | 5.31 ×10$^{-2}$ |
| 0.924 M | 1.84 ×10$^5$ | 0.50 ×10$^3$ | 8.75 ×10$^{-5}$ | 15 ×10$^3$ | 75.0×10$^3$ | 4.96 ×10$^{-2}$ |

Hoerantner et al. [116] employed TRPL and THz spectroscopy techniques, reporting long-range mobilities (μ_long-range) of up to 6.7 cm$^2$/V·s at a photoexcitation density of $10^{15}$ cm$^{-3}$, and revealed a correlation between mobility and crystallite structure (Figure 12). The authors also investigated photoinduced changes in the refractive index (Δn) of the perovskite following excitation. Figure 12a presents the experimentally measured reflectivity variation and the calculated Δn, both showing pronounced changes near the band edge. Below the bandgap, a negative Δn is observed, which becomes more pronounced at lower photon energies and is attributed to band filling and intraband absorption effects. Upon photoexcitation, electrons occupy the lower states of the conduction band, effectively increasing the minimum energy of interband transitions (the band filling effect), thereby widening the apparent bandgap and producing a characteristic peak near the absorption edge (Figure 12a).

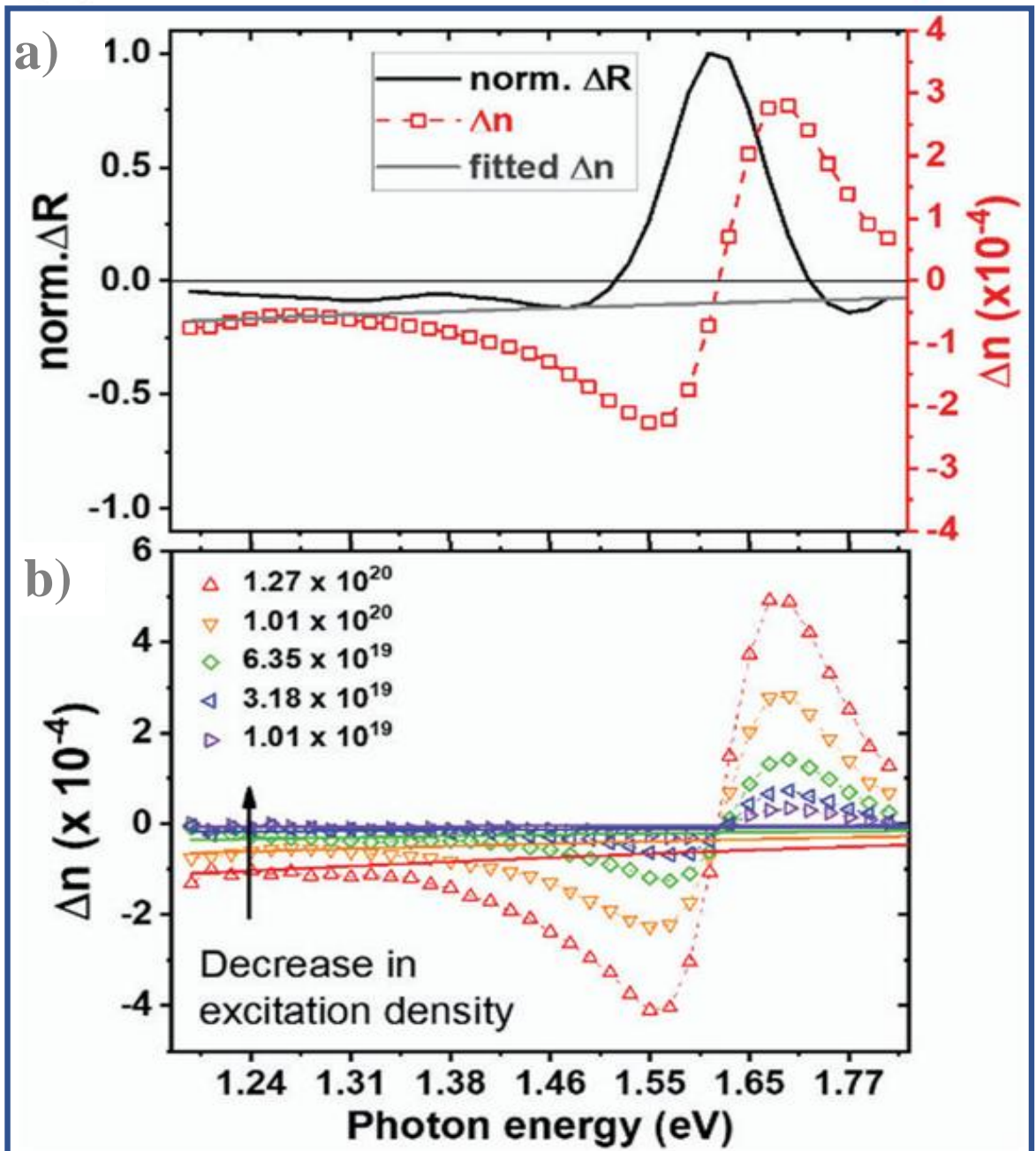

**Figure 12.** Photoinduced changes in the refractive index. (a) Normalized change in reflectance (ΔR) and the corresponding calculated change in refractive index (Δn). (b) The dependence of Δn on the excitation density. Experimental data are shown as symbols, while the model approximation is indicated by solid lines [116].

Free charge carriers also contribute to intraband absorption, known as the plasma effect, which leads to an additional modification of the real part of the refractive index. This behavior is illustrated in Figure 12b and can be accurately described using the Drude model [117]. In the energy range from 1.2 to 1.3 eV, where the contribution from band filling becomes negligible, the observed Δn is predominantly governed by the concentration of free carriers. Within this range, the authors performed a model fitting using equation (3), which is represented by the gray line in Figure 12a [116]. Assuming identical effective masses for electrons and holes ($m_e = m_h = 0.208\,m_e$) and equal carrier densities, they calculated the carrier concentration. The obtained values were consistent with those reported for other direct-bandgap semiconductors such as GaAs. Lee et al. [118] conducted a study in which the addition of KI and $I_2$ to $MAPbI_3$ improved the charge carrier mobility from 0.126 to 0.30 cm$^2$/V·s. The photoluminescence lifetime increased from hundreds of nanoseconds to approximately 1.2 to 1.5 microseconds. These enhancements led to a power conversion efficiency close to 19.4 percent. In Figure 13a, the photoluminescence spectra are compared for untreated films, films treated with KI, and those treated with both KI and $I_2$. The strongest PL signal was observed in the untreated sample, which may be due to a high density of surface and bulk traps. In the doped samples, especially those treated with both additives, the PL intensity decreased, indicating more efficient carrier extraction and reduced recombination losses.

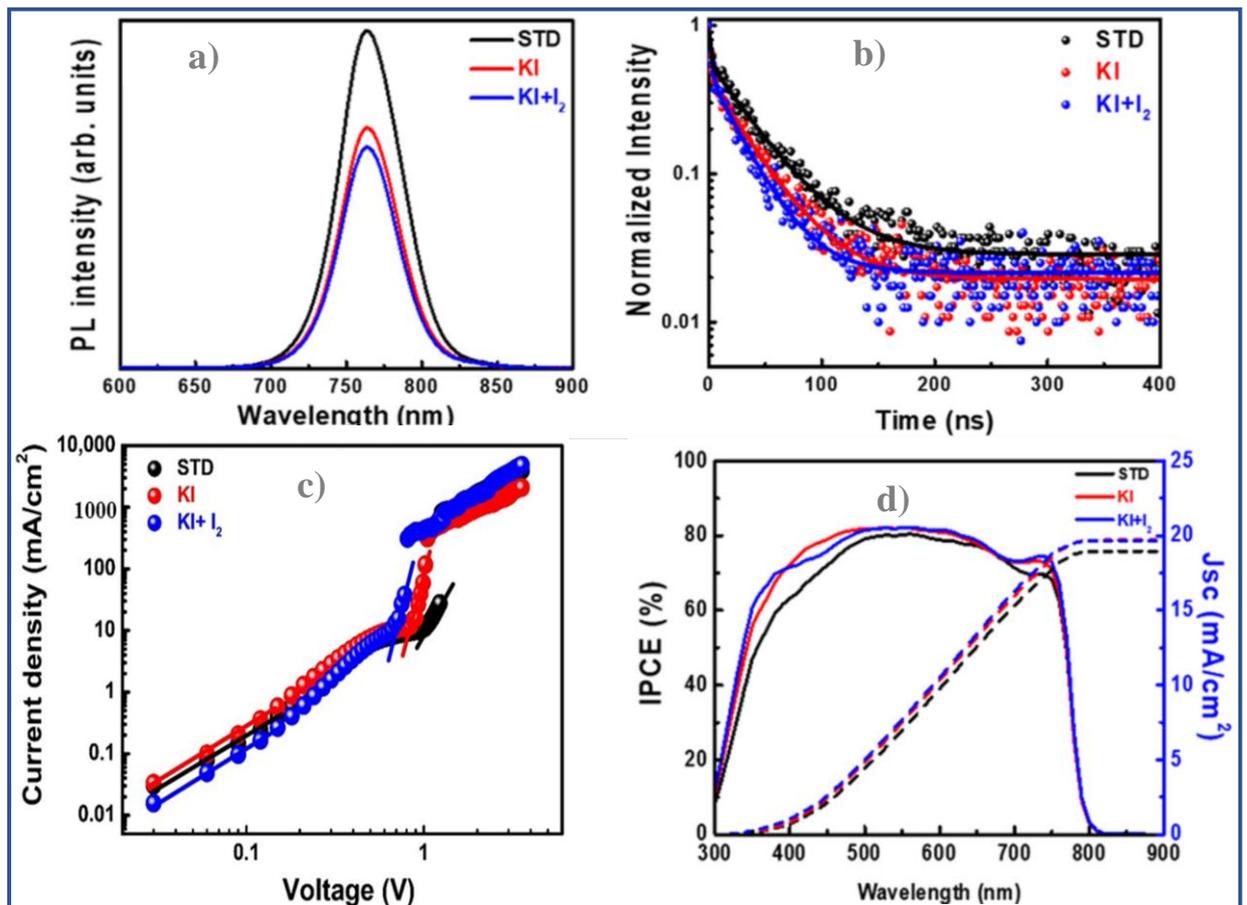

**Figure 13.** The (a) PL and (b) TRPL analyses of perovskite films on TiO2/FTO glass with various optimal conditions (STD, KI (30 mM), and KI (30 mM) + $I_2$ (3 mM)). (c) J–V curves and (d) IPCEs of PSCs with various optimal conditions (pristine, KI (30 mM), and KI (30 mM) + $I_2$ (3 mM)) [118].

The dynamics of photogenerated carrier lifetime presented in Figure 13b confirm this behavior. Doped samples exhibited prolonged lifetimes, especially in the case of KI$^+$ $I_2$, where values reached 1.2 to 1.5 µs. This indicates suppressed recombination and a less defective structure. The TRPL data are in full agreement with the photoluminescence observations and the reduction in hysteresis index. Additional evidence of improved transport quality is provided by the current-voltage characteristics obtained using the SCLC method, as shown in Figure 13c. The sample with $KI^+ I_2$ additive displayed a higher current and steeper slope in the trap-filled limit region compared to the pristine and KI-treated samples. This suggests a lower trap density and enhanced carrier mobility. The transition to a more pronounced quadratic current-voltage behavior further confirms the formation of a uniform, high-quality active layer with improved transport properties. Moreover, better IPCE values were observed across both the short-wavelength and long-wavelength regions (Figure 13d), indicating improved perovskite crystallinity and reduced bulk defects. These improvements are reflected in the enhanced photoelectric performance. In summary, the analysis of Figure 13 clearly demonstrates that the combined $KI^+ I_2$ additive effectively reduces defect density, improves carrier separation and transport, and enhances both stability and efficiency of perovskite solar cells.

Cheng et al. [119] introduced TMDS doping into PCBM at the ETL-perovskite interface. This led to a 26% increase in carrier mobility, a τ_PL of 2.8 µs, and a reduction in $V_{oc}$ losses to 55 mV. These findings highlight the synergy between chemical engineering and charge transport properties. Slimani et al. [120] demonstrated that residual strain reduces both τ and µ, and mechanical relaxation improves performance. After relaxation, mobility increased to approximately 1.8 cm$^2$/V•s and τ_PL reached 1.4 µs, emphasizing the importance of microstructure control. Innovative device architectures incorporating 2D/3D heterostructures represent a promising approach to charge carrier management. Ber-Czerny et al. [121] compared the photophysical response of the 2D perovskite $PEA_2PbI_4$ with the 3D perovskite $FA_{0.9}Cs_{0.1}PbI_3$ using time-resolved photoconductivity measurements. Figure 14a shows photoconductivity decay curves following laser excitation, displaying mono-exponential decay with lifetimes on the order of hundreds of nanoseconds and peak conductivities ranging from $10^{-2}$ to $10^{-4}$ S/cm. These results indicate the presence of long-lived free carriers in both materials. However, simple extrapolation to t = 0 without considering early-time recombination underestimates $\phi\sum\mu$, where $\sum\mu$ is the sum of electron and hole mobilities, and $\phi$ is the fraction of free carriers.

To obtain more accurate mobility values, the authors applied a model accounting for early recombination and Saha equilibrium. Figure 14b presents both corrected and uncorrected values of $\sum\mu$. For the 3D perovskite $FA_{0.9}Cs_{0.1}PbI_3$, the average corrected mobility was approximately 0.8 cm$^2$/V•s, while for the 2D perovskite $PEA_2PbI_4$ it was significantly higher at about 8.0 cm$^2$/V•s. This behavior is attributed to the high exciton binding energy (~230 meV) in the 2D structure and the low fraction of free carriers ($\phi$ ~0.03–0.11), despite their high local mobility. The analysis also revealed that increasing the excitation density led to a decrease in $\phi$ due to enhanced early recombination. No signs of superlinear photoconductivity growth were observed, ruling out trap-filling effects within the tested range.

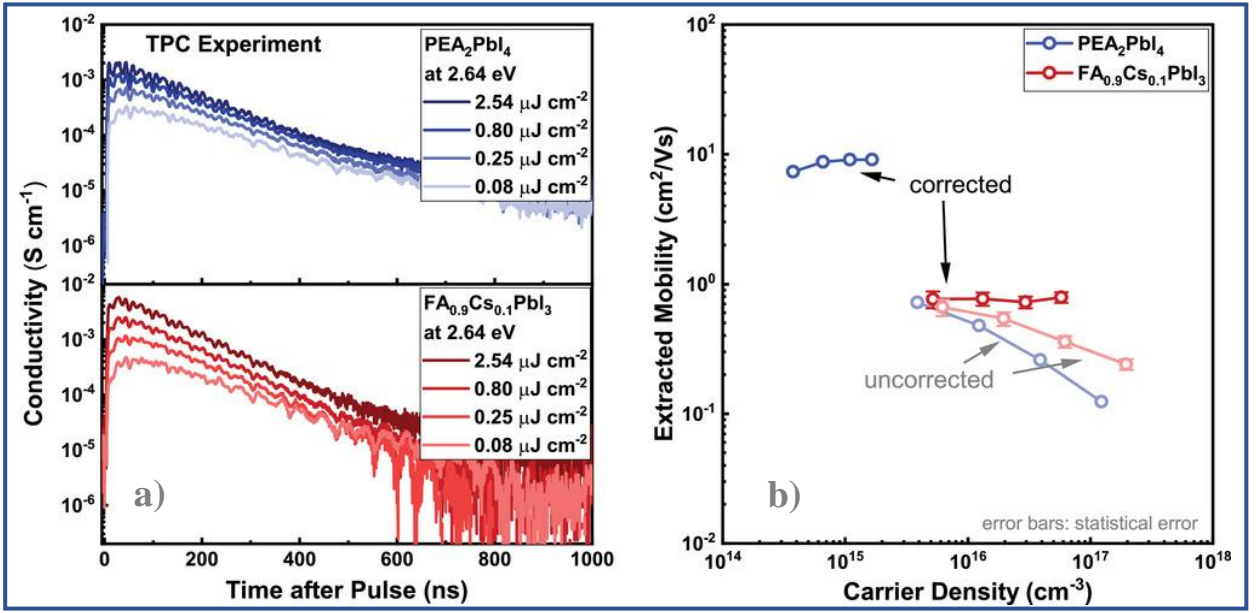

**Figure 14.** (a) Transient photoconductivity decay following excitation by a pulsed 2.64 eV (470 nm) laser at a repetition rate of 10 Hz for $PEA_2PbI_4$ and $FA_{0.9}Cs_{0.1}PbI_3$ under various excitation fluences (represented by color intensity; 0.08–2.54 µJ cm$^{-2}$). (b) Uncorrected $\phi\sum\mu$ and corrected $\sum\mu$ values are shown for both materials, with associated error bars [121].

A substantial contribution to this field has come from multidimensional diagnostic studies combining TRPL, ECPL, TAS, TRMC, and OCVD methods, which have enabled a more comprehensive understanding of recombination pathways [122, 123]. The integrated analysis presented in [124] confirmed that interfacial defects and grain boundaries continue to play a dominant role in limiting both carrier mobility and lifetime, even in highly crystalline materials. Recent findings regarding excitonic and ionic mobility in 2D/3D hybrid perovskite structures reveal that spatial separation and interfacial protection provided by the 2D phase result in longer carrier lifetimes, reaching values between 4 and 6 microseconds. At the same time, the carrier mobility reaches values ranging from 3.2 to 4.5 cm$^2$ V$^{-1}$ s$^{-1}$ [125–127]. The use of PEA$^+$ interlayers supports the formation of an internal electric field, which helps prevent backward diffusion and reduces interfacial recombination. This architecture also suppresses defect ion migration and promotes improved device stability under long-term operation [116].

A notable trend is the use of advanced optical techniques such as transient photoconductivity (TPC) and terahertz (THz) spectroscopy, as demonstrated in the study by Ber-Czerny et al. [121]. That work investigated charge transport anisotropy in 2D and 3D perovskite systems. The analysis of the dependence of carrier mobility on excitation fluence, as well as the estimation of $\sum\mu$ values, indicated that 2D perovskites exhibit high local mobility, even though the density of free carriers is relatively low. This highlights the need for careful interpretation of TRPL results, especially in systems where ultrafast recombination and excitonic states may interfere with the photophysical response [128, 129].

Residual mechanical stress and microstrain-induced defects have also proven to be significant factors affecting carrier transport. Careful control of film deposition parameters and targeted compositional doping help reduce the density of defects that act as nonradiative Shockley–Read–Hall recombination centers [130]. Furthermore, interfacial modification using materials such as PCBM and TMDS has been shown to enhance both stability and photoluminescence lifetimes, which in some cases reach up to 2.8 microseconds. These results demonstrate that the lifetime of charge carriers is influenced not only by the active perovskite layer but also by the nature and quality of the charge transport layers [131].

The application of logarithmic binning methods, as introduced in [114], has significantly improved the detection of photoluminescence decay across a broad dynamic range, from 100 nanoseconds to several milliseconds, while maintaining a low noise level. This technique allows for more accurate detection of slow recombination components, which are often masked during traditional TRPL measurements [132]. Achieving such high sensitivity is essential for evaluating the long-term reliability of devices subjected to extended illumination and thermal cycling. Ion migration remains one of the most critical challenges in perovskite solar cells. According to studies [133–135], the motion of ions such as $MA^+$, $I^-$, and $FA^+$ contributes to the formation of internal electric fields that shift energy levels and lead to a decrease in both carrier lifetime and open-circuit voltage. Additives that limit ionic movement, including large organic cations and interfacial buffer layers, have shown the ability to stabilize charge distribution and device operation. Other investigations [94, 135] emphasize that well-ordered microstructures with minimal grain boundaries and reduced concentrations of Pb and I antisite defects enable diffusion lengths as high as 2 microns and carrier mobilities exceeding $10\,cm^2\,V^{-1}\,s^{-1}$ [136]. These findings reinforce the importance of high film quality in achieving optimal charge transport, regardless of interfacial band alignment.

The integration of strategies such as compositional tuning, charge transport layer engineering, interface modification, and morphology control has enabled power conversion efficiencies above 25 percent while maintaining excellent device stability and mechanical reliability. These results reflect not only advancements in understanding the mechanisms of charge generation, transport, and recombination, but also a readiness for commercial implementation [137–140]. In summary, optimizing the carrier lifetime from 0.15 to 3.3 microseconds and the carrier mobility from 0.12 to $7\,cm^2\,V^{-1}\,s^{-1}$ has become achievable through a combined approach involving chemical additives, structural modification, and rigorous methodology. This has led to devices that reach open-circuit voltages above 1.27 volts, fill factors over 85 percent, and efficiencies greater than 25 percent. These achievements represent not only a milestone in scientific research but also a significant step toward the commercial viability of perovskite solar cells.

## 6. Interface Engineering and Stability Enhancement Strategies in PSCs

The interface between the perovskite absorber and adjacent functional layers, particularly the electron and hole transport layers (ETL and HTL), is critically important for effective charge extraction and long-term operational stability [139, 140]. Engineering these interfaces to minimize trap states, align energy levels, and prevent degradation has become a key strategy in developing next-generation PSCs. One promising approach involves the passivation of defects at the ETL/perovskite interface. Wu and colleagues [141] demonstrated that treating the $SnO_2$ surface with guanidine hydrochloride (GCl) improves the crystallinity of the perovskite layer, reduces trap density, and enhances the air-stability of the device. Figures 15a and 15b show scanning electron microscopy images of a mixed-cation perovskite film spin-coated onto both untreated and GCl-treated $SnO_2$ substrates. In both cases, the films appear compact and uniform without pinholes. However, the average grain size increases from 714 nm on the untreated substrate to 936 nm on the GCl-treated one.

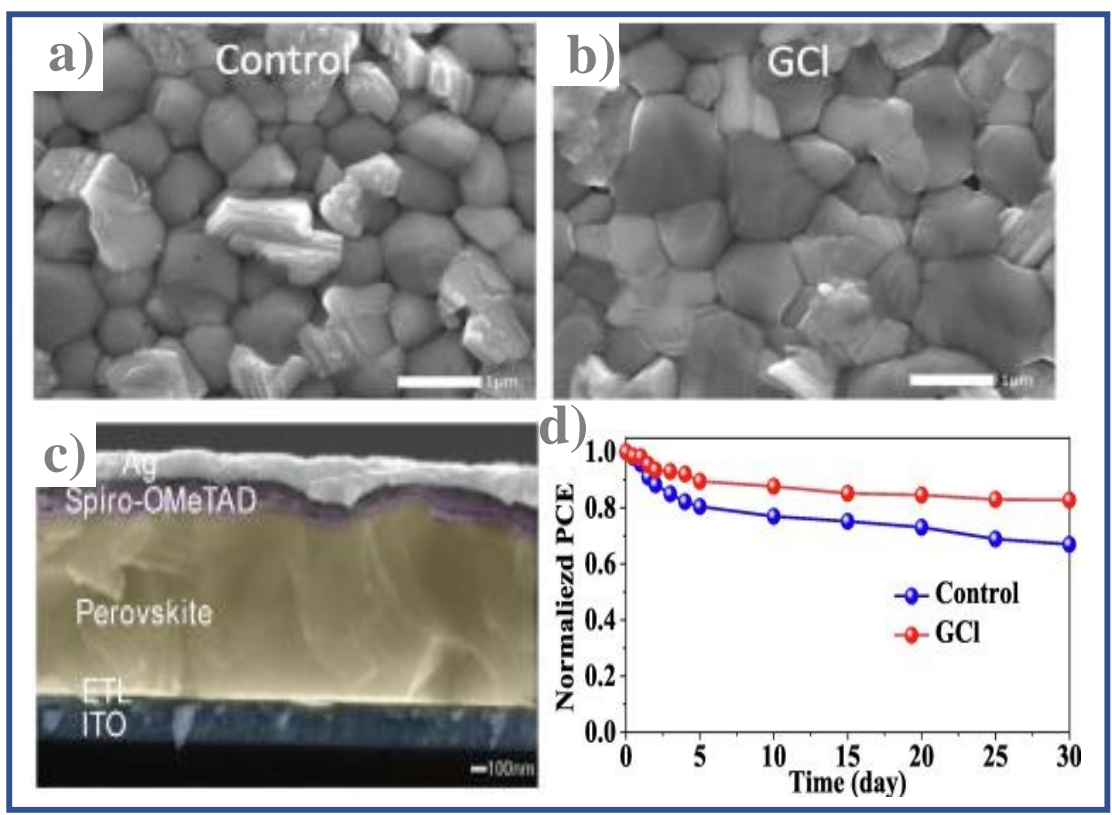

**Figure 15**. Top-view SEM images of perovskite on (a) $SnO_2$ and (b) $SnO_2$/GCl. (c) cross-sectional SEM image of *n-i-p* PSCs structure. (**d**) The normalized efficiency change of the devices with $SnO_2$ and $SnO_2$/GCl as ETLs kept under room temperature in the air [141].

The GCl molecules interact electrostatically with oxygen vacancies in $SnO_2$ and form hydrogen bonds with halide anions in the perovskite layer, creating a strong interfacial connection. This structure helps reduce charge recombination and contributes to improved device performance, as summarized in Table 6. After 30 days of storage without encapsulation, the devices retained 83 percent of their initial power conversion efficiency (Figure 15d). Compared to the untreated $SnO_2$ ETL, the conduction band edge (ECB) of the $SnO_2$/GCl system is better aligned with that of the perovskite absorber, promoting more efficient electron transfer across the interface. A cross-sectional SEM image of the complete cell structure is shown in Figure 15c, where the perovskite layer is approximately 1 micron thick, with large grains extending across the entire thickness of the light-absorbing layer.

In reference [142], the application of ionic self-assembled monolayers (ISAM) and bifunctional molecules is discussed as a means to generate interfacial dipoles, align energy levels, and suppress hysteresis effects. The study also highlights that the incorporation of 4-aminobenzoic acid and benzylamine derivatives enables simultaneous modification of the interfacial energetics at the ETL/perovskite boundary and improves the fill factor of the solar cell [143, 144].

Table 6. Device performance of cells based on $SnO_2$ or $SnO_2$/GCl ETLs [141].

| Samples | $J_{SC}$ (mA cm$^{-2}$) | $V_{OC}$ (V) | FF | PCE (%) |
|---|---|---|---|---|
| Control F | 23.14 | 1.10 | 0.75 | 19.09 |
| Control R | 23.19 | 1.11 | 0.76 | 19.56 |
| GCl F | 23.66 | 1.14 | 0.80 | 21.58 |
| GCl R | 23.71 | 1.14 | 0.80 | 21.63 |

The use of large organic cations, such as phenylethylammonium, butylammonium, and guanidinium, has attracted significant interest due to their ability to form quasi-2D phases at the surface of perovskite films. These cations help reduce surface recombination rates and extend carrier lifetimes [145, 146]. Seshaiah and colleagues [142] also demonstrated that treatment with $PEA^+$ contributes to defect passivation and the formation of low-dimensional structures, while also acting as a barrier against moisture and ion migration. At the HTL/perovskite interface, interface engineering strategies have also proven effective. Wang and co-authors [147] applied poly(triarylamine) derivatives as passivating agents, which significantly reduced nonradiative losses and enabled stable operation of devices for over 1,000 hours. This treatment enhanced hole extraction and contributed to improved energy level alignment [148, 149].

Fullerene-based interfacial modifiers such as PCBM and ICBA are widely employed to passivate trap states and facilitate electron extraction. The authors of [150] reported that such modifications increase photoluminescence quantum efficiency (PLQE) and improve device stability under humid conditions. In other studies [151, 152], trifluoromethylbenzamidine-based modifiers were used to introduce interfacial dipoles that stabilize the internal electric field and suppress recombination. Devices incorporating these modifications maintained stable operation for over 1,000 hours under continuous illumination [152]. Another promising strategy involves the use of nanostructured ETLs with controlled doping. Materials such as Li-doped $SnO_2$ and La-doped $TiO_2$ have shown reduced oxygen-vacancy-related defects and enhanced electron mobility, contributing to thermal stability [153–155]. Furthermore, graded heterostructures within the perovskite itself offer new opportunities for interfacial control. As shown in [156], introducing a bromide-rich compositional gradient layer results in spatial carrier separation and suppressed recombination, effectively doubling the device photostability under sunlight.

Characterization techniques including time-resolved photoluminescence (TRPL), transient photocurrent (TPC), and electrochemical impedance spectroscopy (EIS) have confirmed the effectiveness of interfacial engineering. These methods reveal increased carrier lifetimes, reduced recombination resistance, and minimized hysteresis. In particular, the hysteresis index decreased from 2.44% to 0.23% after GCl treatment, and TPV measurements showed a notable increase in recombination lifetime [141]. Studies indicate that devices with well-engineered interfaces retain 80 to 90 percent of their initial efficiency after several weeks of exposure to environmental conditions [157, 158]. This suggests that the primary degradation pathways are often associated with interfacial defects, such as uncoordinated lead or halide atoms, which are especially vulnerable to moisture and oxygen. These findings collectively underscore the critical importance of chemical passivation, energy level alignment, and precise morphological and structural control at interfaces. The continued development and implementation of advanced interfacial engineering strategies remains central to improving the efficiency and stability of perovskite solar cells.

## 7. PSC Industrialization: Challenges and Prospects

Despite the rapid progress of perovskite solar cell (PSC) technology over the past decade, several key challenges still hinder widespread deployment. However, a review of recent scientific developments indicates that through advancements in materials science, interface engineering, device architecture, and environmental compatibility, a sustainable path to commercialization can be achieved. One of the most critical directions involves scaling laboratory-scale cells to large-area modules. While single-junction PSCs have achieved efficiencies of up to 25.7% and tandem PSC–Si architectures exceed 31% in laboratory settings [159], scaling up to module areas of 20–24 cm$^2$ often results in reduced efficiencies of around 21–23% [160]. This decline is attributed to challenges in maintaining uniform layer deposition, higher defect densities, and degradation over larger surface areas. To address these issues, new deposition techniques such as slot-die coating, blade coating, and inkjet printing have been developed [160]. These methods support compatibility with roll-to-roll manufacturing and help reduce production costs.

Nevertheless, long-term operational stability remains one of the most pressing concerns. To meet the IEC 61215 standard and ensure lifetimes of 25 to 30 years, issues such as ion migration, organic component degradation, and interfacial deterioration must be addressed [161, 162]. Effective encapsulation technologies and the adoption of more stable inorganic materials, including $CsPbI_3$ and $FAPbI_3$ with additives such as carbazole, have enabled stability of up to 500 hours under 85$^o$C and 85% relative humidity with minimal efficiency loss [161, 163]. Lead toxicity is another major issue drawing attention from both the scientific community and regulatory bodies. Module failure can result in the release of $Pb^{2+}$ ions, posing environmental risks [164]. Two primary strategies have been proposed: physical and chemical encapsulation [164], and the development of lead-free PSCs based on elements such as Sn, Bi, Sb, Mn, and Ge [165]. While Sn-based PSCs have surpassed 15% efficiency, they still face challenges related to $Sn^{2+}$ oxidation and elevated defect densities [27, 166, 167].

Alongside improved reliability and environmental safety, there is growing interest in specialized applications of PSCs, including integration into building elements (BIPV), portable electronics, space technologies, and the Internet of Things (IoT). Each application presents unique requirements. For BIPV, transparency and color stability are essential [168]. In space applications, UV resistance and thermal cycling tolerance are crucial [168]. For IoT systems, high sensitivity under low-light conditions and an optimal bandgap near 1.9 eV are desired [168, 169]. The economic aspect also plays a vital role. Levelized cost of electricity (LCOE) analyses estimate that with a 25-year lifespan, PSC-based systems can achieve an LCOE of 0.0348 USD/kWh, which is significantly lower than conventional silicon modules at 0.0550 USD/kWh [170, 171]. The development of tandem architectures and low-cost production methods is expected to drive costs even further down. Additionally, efforts are underway to enable recycling of PSC components at end-of-life. Recovering glass substrates, noble metal electrodes, and lead-containing compounds offers a means of cost reduction and reduced environmental impact [172, 173].

Integration of PSCs with other technological platforms such as energy storage systems, sensors, and flexible electronics is also gaining traction. These solutions are especially relevant for wearable devices and autonomous power systems. However, they require highly elastic conductive substrates and stable organic–inorganic interfaces [168]. Recent advances in stability enhancement, toxicity mitigation, scalability, and broadened applicability bring PSCs closer to commercial readiness. Achieving this goal will require collaboration among materials scientists, engineers, and environmental specialists. At the current pace of innovation, PSCs are poised to become a foundational technology for future energy systems that are efficient, low-cost, and suitable for widespread deployment across sectors such as energy, construction, defense, and consumer electronics.

**Conclusion**

Over the past decade, perovskite solar cells have evolved from a laboratory curiosity into one of the most promising candidates for next-generation photovoltaics. However, commercial adoption requires more than high power conversion efficiency. Long-term stability, environmental compatibility, and scalable manufacturing must also be ensured. This review has demonstrated that comprehensive control over charge carriers, including prolonged lifetimes, enhanced mobility, trap passivation, and optimized interfacial engineering, plays a central role in overcoming current limitations in PSCs. The article has highlighted how synergies among film morphology, interfacial strategies, and device architecture can effectively reduce nonradiative recombination and improve charge extraction.

Particular attention was paid to stability-enhancing strategies such as encapsulation, defect-tolerant designs, barrier layer introduction, and the use of 2D/3D hybrid structures. These approaches contribute to more reliable operation under real-world conditions. Nonetheless, several challenges remain. These include maintaining film uniformity during scaling, mitigating lead toxicity, and implementing effective lead-free alternatives. Looking forward, interdisciplinary efforts that combine materials science, device physics, quantum chemical modeling, and machine learning are expected to shape the next phase of PSC development. The combination of technological innovation and strategic research is likely to establish PSCs as a key technology in the global transition toward sustainable energy systems.

**FUNDING**
This work was supported by the Interstate Fund for Humanitarian Cooperation of the CIS Member States through the scientific project funded by the International Nanotechnology Innovation Center of the CIS (grant no. 25-113), and by the International Science and Technology Center (grant no. TJ-2726).

**CONFLICT OF INTEREST**
The author declared no conflict of interest.